\documentclass[]{aa}

\usepackage{graphicx}
\usepackage{txfonts}
\usepackage{natbib}
\bibpunct{(}{)}{;}{a}{}{,}

\def\f0{$f_0({\bf v})$}

\def\ROLR{R_{\rm OLR}}

\def\ds{\displaystyle}
\def\kms{\ km s$^{-1}$}
\def\kmskpc{\ km s$^{-1}$ kpc$^{-1}$}

\def\<#1>{\langle#1\rangle}
\def\pcite#1{[ref]}

\begin{document}
   \title{Phase Space Structure in the Solar Neighbourhood}
   \subtitle{}
   \author{Dalia Chakrabarty,
          \inst{1}
          }
   \offprints{Dalia Chakrabarty}
   \institute{School of Physics $\&$ Astronomy,
              University of Nottingham, 
              Nottingham NG7 2RD, U.K.
              \email{dalia.chakrbarty$@$nottingham.ac.uk}
             }

   \date{\today}

\abstract {} {To examine the idea that dynamical parameters can be
estimated by identifying locations in the solar neighbourhood where
velocity distributions recovered from test particle simulations, match
the observed local distribution. Here, the dynamical influence of both
the Galactic bar and the outer spiral pattern are taken into account.}
{The Milky Way disc is stirred by analytical potentials that are
chosen to represent the two perturbations, the ratio of pattern speeds
of which is explored, rather than held constant. The velocity
structure of the final configuration is presented as heliocentric
velocity distributions at different locations. These model velocity
distributions are compared to the observed distribution in terms of a
goodness-of-fit parameter that has been formulated here. We monitor
the spatial distribution of the maximal value of this goodness-of-fit
parameter, for a given simulation, in order to constrain the solar
position from this model. Efficiency of a model is based on a study of
this distribution as well as on other independent dynamical
considerations.}  {We reject the bar only and spiral only models and
arrive at the following bar parameters from the bar+spiral
simulations: bar pattern speed of 57.4$^{+2.8}_{-3.3}$\kmskpc and a
bar angle in [0$^\circ$, 30$^\circ$], where the error bands are
$\pm$1-$\sigma$. However, extracting information in this way is no
longer viable when the dynamical influence of the spiral pattern does
not succumb to that of the bar; an explanation for this is
offered. Orbital analysis indicates that even though the basic
bimodality in the local velocity distribution can be attributed to
scattering off the Outer Lindblad Resonance of the bar, it is the
interaction of irregular orbits and orbits of other resonant families,
that is responsible for the other moving groups; it is realised that
such interaction increases with the warmth of the background disk.}
{}


   \maketitle

\section{Introduction}
\noindent
Historically, stars in the solar neighbourhood were considered
to exhibit purely random peculiar motion; guided by this notion, the
motion of the Sun was calculated from chosen stellar samples.  This
idea was challenged by Jacob Kapteyn at the beginning of the last
century when he observed preferential stellar motion in two favoured
directions. Within the next few decades, the situation was realised to
be more complicated, though the dominance of the two Kapteyn streams
(Stream I $\&$ II) was established. The apex of the motion of the
stars in Stream I was observed to correspond to the convergent point
of the proper motions of the Hyades cluster \citep{olina} while that
of the members in Stream II almost coincides with the convergent point
of proper motions of the Sirius supercluster. Thus Stream I is also
referred to as the Hyades stream, while Stream II is referred to as
the Sirius stream.  As observations improved, more moving groups were
observed in the solar neighbourhood, such as the Hercules, Coma
Berenices and Pleiades streams, \citep{fux}.

In the past, work has been done to understand the origin of the moving
groups in the solar neighbourhood, as handiwork of the bar or
transient spirals \citep{walterolr, fux_aa, tremainewu, quillen,
quillenminchev}. But no such exercise took into account the joint
effect of these two disc structures, while scanning through a possible
range of the ratio of their pattern speeds. This ratio is foreseen to
have serious dynamical influence in sculpting the local phase
space. Such modelling is included in this paper. 

Also, while work has been done to estimate the bar parameters (pattern
speed and bar angle) from a comparison of simulated velocity
distributions $f_s$ and the observed velocity distribution $f_o$
\citep{walterolr}, a rigorously quantified comparison of the same has
not been carried out. This obviously hinders the possibility of
scanning through an assortment of models and also renders such
parameter estimates subjective. In this paper, a statistical formalism
is presented that allows for such objective comparison. In fact, this
methodology points to the fallibility of attempting to extract
dynamical parameters from such comparison, in cases when the influence
of the spiral is strong.

\begin{figure}
\centering
\includegraphics{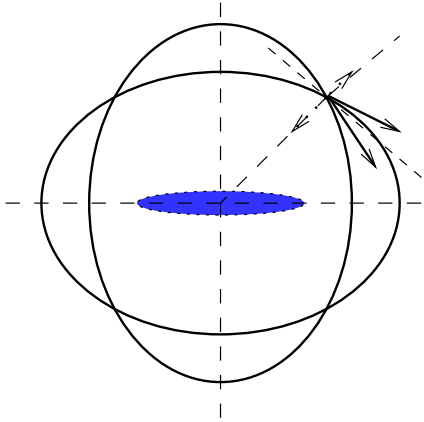}
\caption{The OLR location is marked by orbits of opposite orientations
on either side of the resonance; orbits are aligned with the major axis of
the bar just outside the resonance while just inside, circular orbits
are distorted into anti-aligned shapes. The local velocity distribution,
as viewed by an observer sitting at the junction of these two different
orbits, is bimodal. The bar is shown by the solid ellipse at the centre.}
\label{fig:kalnajs}
\end{figure}

Additionally, multiple suggestions appear to have been put forward
toward a dynamical origin for the moving groups. In \citet{agris}, it
was argued that if the Outer Lindblad Resonance due to the central bar
($OLR_b$) in our Galaxy corresponded roughly to the solar radius, then
an observer at the Sun would be close to the point of intersection of
the aligned and anti-aligned orbits that lie on either side of the
resonance location and identify the local stellar distribution as
bimodal, (Fig~\ref{fig:kalnajs}). This bimodality was also picked up
by \citet{palous} in a paper that reported the distribution of a
sample of A-type stars in the solar neighbourhood.  In
\citet{walterolr}, results of simulations done (by backward
integration) with a central bar imposed on a disk, were reported; it
was concluded that the bar angle lies in the range
[$10^{\circ},70^{\circ}$] and the bar pattern speed is
$53\pm3$\kmskpc. This value falls slightly short of $59\pm2$\kmskpc that is suggested
for the bar pattern speed in the hydrodynamical work by \citet{peter}
but falls within the range suggested in an
application of the Tremaine-Weinberg method to a sample of OH/IR stars
in our Galaxy by \citet{debattista} - $59\pm5$\kmskpc with a possible
systematic error of 10\kms.

\citet{fux} discusses the results of a high-resolution N-body
simulation aimed at investigating barred models of the Galaxy. The
local velocity distribution function was smoothed by the adaptive
kernel smoothing technique discussed in \citet{skuljan}. It is to be
noted that this sample is biased since the used radial velocities are
for the high proper motion stars only \citep{binney97, skuljan,
fux_aa}. Nonetheless, stellar streams are distinctly visible in this
distribution as well as in the model distributions recovered by
Fux. According to Fux, $\ROLR=7.7$ kpc and the bar angle is
$25^{\circ}$. Figure~\ref{fig:fuxuv} represents the local velocity
distribution, as estimated in \citet{fux}.  The velocity data
corresponding to this figure and the smoothing algorithm used to
deduce the contour plot presented therein, were very kindly supplied
by R.Fux.

The observed bimodality in the velocity distribution in the solar
neighbourhood has been explained in \citet{fux} along the same lines
as in \citet{raboud}. The Jacobi integral, in the rotating frame of
the bar, is stationary at five Lagrange points on a zero-velocity
surface; these five characteristic values of the integral correspond
to two maxima, one minima and the two saddle points, in the effective
potential. If the Jacobi energy of a star exceeds the value of the
effective potential that characterises the saddle points, then the
star is free to cross corotation and escape to infinity, (in
principle). Following \citet{fux}, such stars are said to be on
``hot'' orbits. The Hercules stream is concluded to be due to stars on
such orbits. The stability of the orbits is dealt with in details in
\citet{fux_aa}; in this work, the Hercules stream is proposed to be
due to an overdensity of chaotic orbits at spatial locations that are
near or just outside the OLR.

\citet{tremainewu} propose the heating of the disk by strong transient
spiral waves as the source of the moving groups that are evident in
the local phase space structure. They also suggest that this mechanism
can explain the observation that ages of the stars in the same moving
group vary over a wide range. \citet{famaey05} agree that this wide
variation of ages within the same moving group can be understood as
due to the migration of stars from an original galactocentric
location, brought about by a transient spiral wave and that the
streams are dynamical in origin and not derivatives of irregularities
in the star formation rate.

\citet{quillen} deals with the case of the perturbation due to a bar
and a spiral pattern, as was investigated in Paper~I. From the mapping
of the phase space via Poincare maps, \citet{quillen} realised that
the quasi-periodic orbits that support the bar and the spiral
structure, are disrupted around the OLR of the bar when the solar
radius just exceeds the location of the ILR of the spiral
pattern. \citet{quillenminchev} attribute the splitting of the
Hyades-Pleiades group to spirality.


\citet{paper1} (hereafter Paper~I) reported a series of direct two
dimensional test particle simulations in which the outer parts of
different disk configurations (cold Mestel and warm quasi-exponential)
were stirred by non-axisymmetric perturbations due to the bar alone, a
spiral pattern alone, and the bar and the spiral acting in concert.
The result of the perturbation was gauged via the distributions in the
space of the heliocentric radial velocities ($U$) and tangential
velocities ($V$), disk heating and spatial re-arrangement of the stars
in the disc. The effects of growing and subsequently dissolving a
perturbation were also looked into.

While in Paper~I we attempted to understand the general perturbative
effects of non-axisymmetric features, in this paper, such an
investigation is configured to model the effects of the Galactic bar
and outer spiral pattern on the phase space distribution in the
vicinity of the Sun.



The paper has been organised as follows. In the following section, we
discuss the methodology that is used here. In
Section~\ref{sec:background}, we elucidate the basic parameters used
in our models and introduce the goodness-of-fit statistic that we use.
In Section~\ref{sec:distr}, velocity distributions obtained from our
simulations have been presented at various locations, for the
different runs that have been performed. An analysis of some orbits
responsible for the main structures observed in the velocity space is
presented in Section~\ref{sec:orbits}. We then proceed to discuss
some of the issues that the results alluded to, in
Section~\ref{sec:discussions}. The paper is rounded up with a short
summary of the main results.

\section{Method}
\label{sec:method}
\begin{figure}
\centering
\includegraphics[width=8cm]{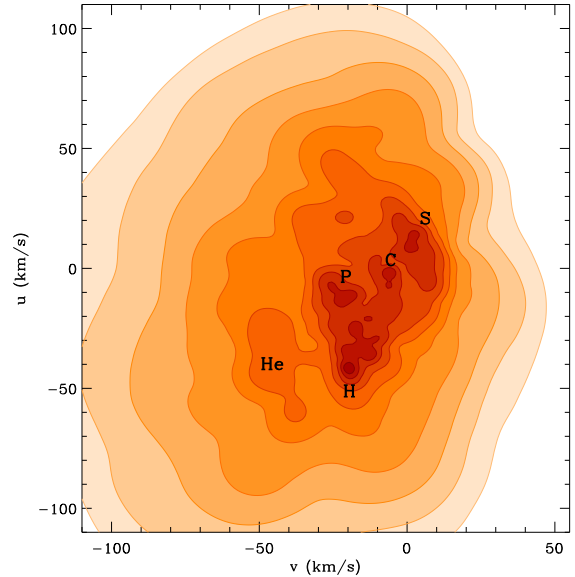}
\caption{The solar neighbourhood distribution diagram produced with
the velocity data and smoothing code, shared with us by Fux. The
distribution above differs from Figure~1 in Fux (1999) in the
definition of the direction of positive $U$, (towards Galactic centre
in the figure above). $V$ is positive in the sense of Galactic
rotation. The above distribution is built from the transverse
velocities of {\it Hipparcos} single stars with distance $d < 100$pc,
dispersion of parallax $\sigma(\pi) < 0.1\pi$ from the Hipparcos
Catalogue and with radial velocities of the 3481 stars in the {\it
Hipparcos} Input Catalogue. Logarithmic contours have been used.}
\label{fig:fuxuv}
\end{figure}

\noindent
In this paper, we undertake an exercise similar to that in Paper~I,
but configure the structures to a model for the Milky Way disc. The
initial positions of the stars are extracted from a model for the disk
(discussed below) and the orbits are numerically integrated in the
potential of the background disk, on which is imposed the analytical
potential of the model Galactic bar and/or the outer spiral
pattern. The spatial band under investigation is put on a regular
polar grid. At each $R-\phi$ cell in this grid, the resulting orbits
are put on a regular Cartesian $U-V$ grid. The velocity distribution
recovered at an $R-\phi$ cell is compared to the observed local $U-V$
distribution (Figure~\ref{fig:fuxuv}). If the comparison is
favourable, the corresponding $R-\phi$ location is branded ``good''. A
rigorous statistical formalism to evaluate the quality of this
comparison is presented below. Thus, a model velocity distribution is
``good'' if it corresponds to a high value of the goodness of fit
index.

The spatial distribution of these ``good'' locations are then used to
constrain the solar position, which is subsequently implemented to
estimate relevant dynamical parameters, such as the bar angle and the
bar pattern speed. (In our scale free disk, we express all lengths in
units of the corotation radius $R_{\rm CR}$ of the bar. Thus,
identifying the solar radius enables us to scale all lengths to real
units; given that the bar pattern speed determines $R_{\rm CR}$, the
scaling implies constraining the bar pattern speed.)

\section{Simulation Background}
\label{sec:background}
\noindent
Here we discuss some of the salient features of our simulations.

\subsection{Models}
\label{sec:models}
\noindent
The models that we have used in our simulations are typically
characterised by two classes of parameters: disk and perturber. An
important feature of the perturbing potential is the strength
parameter; this is the ratio of the field due to the imposed perturber
at a chosen radius, (namely the OLR due to an $m=2$ perturbation in a
Mestel potential) to the field due to the background
disc. Additionally, the simulations that include the spiral pattern,
are distinguished from each other in terms of the choice of the
ratio of the pattern speed of the spiral to the bar. The other
parameters of the spiral pattern, such as the number of arms and pitch
angle are held a constant.

The observational milieu that inspires the choice of the values of
these parameters is discussed below.
\begin{enumerate}
\item The Galactic disk was modelled to have a uniform rotation curve,
\citep{leon} and an almost exactly exponential surface density profile
with a scale length of about 0.9$R_{\rm CR}$. This was ensured by
characterising the disk with the {\it doubly cut-out equilibrium
stellar distribution function}, discussed in \citet{wynjenny,
wynpower} and used extensively in Paper~I. At the end of each
simulation, the radial and azimuthal velocity dispersions ($\sigma_U$
and $\sigma_V$), and the vertex deviation ($l_v$) are noted at
different locations within the annulus that we study. These quantities
should reflect observations on a typical eclectic sample of stars in
the solar neighbourhood (Table~10.2 in \citep{BM}). The values of
$\sigma_U$, $\sigma_V$ and $l_v$ are included in
Table~\ref{tab:results}, which represents the results from the runs
performed with the different models. Thus, we strive to ensure that
the state of the background disk abides by relevant observations. On
the other hand, the state of the disk at time $t=0$ may have been very
different; after all, disk characteristics prior to the growth of the
Galactic bar (and/or outer spiral) are not known to us. Thus, any model
that offers the correct final configuration, suffices. The $\sigma_U$
of the model disk at $t=0$ is about 21 kms$^{-1}$ while the value of
the Oort ratio in the initial disk is about 0.67.

\item This equilibrium model is perturbed either by a quadrupolar bar
or a logarithmic outer spiral pattern, or simultaneously by both these
features. Thus, in the frame that is stationary with the bar, the bar
potential is (as given by Equation~10 in Paper~I):
\begin{equation}
\Psi_{\rm bar} = \epsilon\displaystyle{\frac{\cos(2\phi)}{R^3}}
\end{equation}
Now \citet{walter99} suggests that at the solar radius ($R_\odot$),
the amplitude of $\Psi_{\rm bar}$ is about 0.036$v_0^2R_{\rm
CR}^3/R_\odot^3$, where $v_0$ is the amplitude of the rotation curve
at the solar circle. Now, as in Paper~I, we work in a scale free
Mestel disk, in which, {\it $v_0$=1 and $R_{\rm CR}$=1}, (this $R_{\rm
CR}$ follows from the setting of the bar pattern speed to
unity). Connecting our strength parameter to the strength used by
\citet{walter99}, we get: $\epsilon\approx$0.036$v_0^2R_{\rm
CR}^3$. Now \citet{fux_aa} works with bars that are double in
strength. We choose to adopt the middle path by working with a bar
that has an average of the strengths implemented by \citet{fux_aa} and
\citet{walter99}. This implies that the ratio of the gravitational
field of the bar to that of the background disk is about 3.6$\%$, at the
OLR.

\item The outer spiral pattern is chosen to have 4-arms and a pitch
angle of $15^\circ$. \citet{vallee02} provides a comprehensive review
of the parameters pertinent to the Galactic spiral pattern.  This
review suggests that the pitch angle lies in the range of 6$^\circ$ to
17$^\circ$, with the mean around 12$^\circ$. We work with a pitch
angle of 15$^\circ$, more along the lines of \citet{johnston01}, who
also suggest a 4-armed pattern. These many arms are compatible with
the best-fitting (``standard'') model of the Galactic spiral pattern
in the work by \citet{bissantz}. \citet{blockpuerari} looked at the
$K$-band spiral structure of a sample of 19 galaxies and concluded
that the fractional amplitude with respect to the background disk of
an $m$=2 spiral ranges from 0.03 to 0.5, with the median at 0.1. Now,
the amplitude of an $m$-armed spiral pattern goes as $1/m$
(\citep{tremainewu}). Thus, the range allowed for the fractional
amplitude of the 4-armed spiral, according to \citet{blockpuerari} is
0.015 to 0.25, with a median at 0.05. We choose to work with a 4-armed
spiral that has a fractional amplitude of 0.036. Our spiral strength
is therefore on the weaker side.

\item When the bar and the spiral pattern are imposed on the disk
together, the chosen amplitude for the perturbation field, (at
$OLR_b$), is about $4.9\%$ of the field of the background disk, in
which, the contributions of the bar to the spiral is in the ratio of
0.043:0.023.

\item As discussed in Paper~I, the perturbation is imposed after the
stars had experienced the axisymmetric Mestel potential for
80$\tau_b$/2$\pi$, where $\tau_b$ is the time taken for one bar
revolution. The perturbation was then allowed to grow adiabatically to
its maximum strength. By this we mean that the growth time is much
larger than the dynamical time of the perturber. We chose the growth
times of the bar and the spiral pattern to be equal
(40$\tau_b$/2$\pi$). Orbits were recorded once the perturber strength
had saturated to its maximum.

\item In the disk, a radial band is examined for the effects of the
perturbations. The inner edge of this annulus is at $OLR_b$ and it
extends outwards; specifically, radius ranges from $R/R_{\rm CR} =
1.7$ to $R/R_{\rm CR}=2.3$. At the end of the simulation, the orbital
position coordinates ($R$ and azimuth $\phi$, where $\phi$=0 is along
the bar major axis) are placed on a regular $R-\phi$ grid and the
velocity distributions of the stars lying in each $R-\phi$ cell is
sought. In our simulations the extent of each radial cell is $R/R_{\rm
CR}=0.025$. Therefore, any radial location, in units of $R_{\rm CR}$,
is expressed with multiple significant figures. The azimuthal range
under study is [$0^\circ, 360^\circ$] and any two azimuthal cells are
$10^\circ$ apart. In the presentations of the results, we limit
ourselves to the first quadrant in $\phi$ only.

\item The Hercules stream consists of stars at large negative
tangential velocities. Such stars are on prograde orbits and must
therefore have their guiding centres inside the solar circle. Thus,
the way to boost the Hercules stream is to ensure that more stars from
the inner disk are allowed to enter the annulus being studied. This
can be controlled via the choice of the pattern speed of the spiral
that is used in the simulation as is explained in the next
paragraph. Of course, the surface density of the background disk will
also contribute to this, but this is fixed since we work with a fixed
background configuration (that has been ascribed an exponential
profile).

\item We chose to work with a pattern speed for the spiral pattern
that is distinct from that of the bar. \citet{rautiainen} have
suggested N-body models in which the inner spiral rotates with the bar
while the outer spiral is decoupled from it. \citet{bissantz} have
picked up on this theme to explore gas dynamical models of the galaxy,
in which the bar and the outer spiral rotate with pattern speeds of
about 60kms$^{-1}$kpc$^{-1}$ and 20kms$^{-1}$kpc$^{-1}$
respectively. \citet{melnik} too advocates a similar picture, with the
Cygnus arm as the link between the inner (faster of the two patterns)
and the outer spiral pattern. In fact, \citet{melnik} constrains the
angular speed of the outer spiral pattern ($\Omega_{\rm sp}$) from the top
by requiring that the Perseus arm be inside the corotation due to this
pattern, i.e. $\Omega_{\rm sp}$ $<$ 25kms$^{-1}$kpc$^{-1}$. We take our
cue from this suggestion and choose to work with three pattern speeds
for the outer spiral: 18kms$^{-1}$kpc$^{-1}$, 21kms$^{-1}$kpc$^{-1}$
and 25kms$^{-1}$kpc$^{-1}$. When the 4:1 ILR due to our 4-armed spiral
pattern ($ILR_s$) is constrained to approximately coincide with the
location of $OLR_b$, it implies that $\Omega_{\rm bar}:\Omega_{\rm
sp}=55:21$. In the other two cases, $ILR_s$ lies well inside the
$OLR_b$ ($\Omega_{\rm bar}:\Omega_{\rm
sp}=55:25\:\Longrightarrow\:R/R_{\rm CR}\approx$ 1.42) while in the
other, $ILR_s$ is beyond $OLR_b$ and is located at $R/R_{\rm
CR}\approx$ 1.97, ($\Omega_{\rm bar}:\Omega_{\rm sp}$=55:18). This
radial location sits almost in the middle of the annular region of the
disk that we investigate in our work. Thus in the first case, stars
would be pushed into the radial range under investigation, since the
ILR due to a spiral pattern is an ``emitter'' of angular momentum
\citep{donaldagris}. For this same reason, in the second case, the
effect of the spiral pattern would be to deplete the immediate
neighbourhood of the ILR, i.e. a part of the radial range under
investigation. Thus in the light of the previous paragraph, this
implies that when $ILR_s$ is chosen to lie inside the inner edge of
the radial zone of investigation, the Hercules stream will be more
populous than when the ILR is anywhere inside this zone.
\end{enumerate}

\subsection{Comparison}
\noindent
The solar neighbourhood $U-V$ velocity distribution $f_o$ is generated by
\citet{fux} from the observed data by implementing the adaptive kernel
algorithm suggested by \citet{skuljan} (Figure~\ref{fig:fuxuv}). The
simulated velocity distributions $f_s$ that we obtain at different observer
locations, are similarly smoothed and then compared against the
observed distribution.

To ensure that this comparison is not affected by any extraneous
factors such as differences in the generation or representation of the
(simulated or observed) distributions from the respective data sets,
it was imperative that the model data be smoothed in the same way as
the observed data. This was done by using the same adaptive kernel
smoothing routine that Fux used.

The comparison between $f_s$ and $f_o$ is carried out in terms of a
rigorous ``goodness-of-fit'' statistic (Section~\ref{sec:stats}). Here
are a few points to remember in regard to these comparisons.
\begin{itemize}

\item In making the comparison at any given $R-\phi$ address, we focus
upon the 5 moving groups that have been marked in
Figure~\ref{fig:fuxuv}- $f_s$ is identified as ``good'' if the
locations of all the 5 groups in the $U-V$ plane match the $U-V$
coordinates of the corresponding groups in $f_o$.

\item The match is considered acceptable within error bars that are
given by the solar peculiar velocities
($\vert{V_\odot}\vert\approx$5\kms and
$\vert{U_\odot}\vert\approx$10\kms). It needs to be emphasised that at
any given $R,\phi$ location, the sub $U_\odot-V_\odot$ offset in
velocities that is allowed for a match to be acceptable, is {\it
unique for all} 5 relevant moving groups. In other words, at any
location, we only allow for one translation of the full simulated
distribution along the $U$-axis (by a maximum of 10\kms) and one along
the $V$-axis (by a maximum of 5\kms).

\item It is possible that there is a peak in $f_s$, at a certain point
in the $U-V$ plane but $f_o$ is featureless at this point. Such extra
clumpiness in $f_s$ can be understood as the effect of the following:
\begin{enumerate}
\item It is possible that the observations underestimate the degree of
clumpiness in the local velocity distribution due to shortcomings in
the data, such as measurement errors.
\item Our assumption of smooth and slowly varying potentials may not
be valid; rapid and episodic changes in the potential are possible.
\item Scattering processes are omitted in the simulations, (such as
scattering off molecular clouds and complexes), and these would
contribute towards smoothing the modelled velocity distributions.
\item Perhaps interaction with the live triaxial halo (not included
herein) might lead to smoothening of $f_s$.
\end{enumerate}

Thus, if there is an extra clump in a recovered $f_s$, which does not
correspond to a feature in $f_o$, the discrepancy is not considered to
be the basis for rejecting this simulated distribution. On the other
hand, if there is a peak in $f_o$, which does not have a counterpart
in the model $U-V$ diagram, then such an $f_s$ is rejected. This does
imply that we are testing the hypothesis that all the 5 stellar
streams in the solar neighbourhood can be attributed to the dynamical
influence of the bar and/or the outer spiral. This may indeed not be
realistic and other effects might be relevant. Nonetheless, here we
attempt to demonstrate the efficacy of the non-axisymmetric perturbers
in the generation of the streams.

\end{itemize}

\begin{figure*}
\centering
\includegraphics[width=12cm]{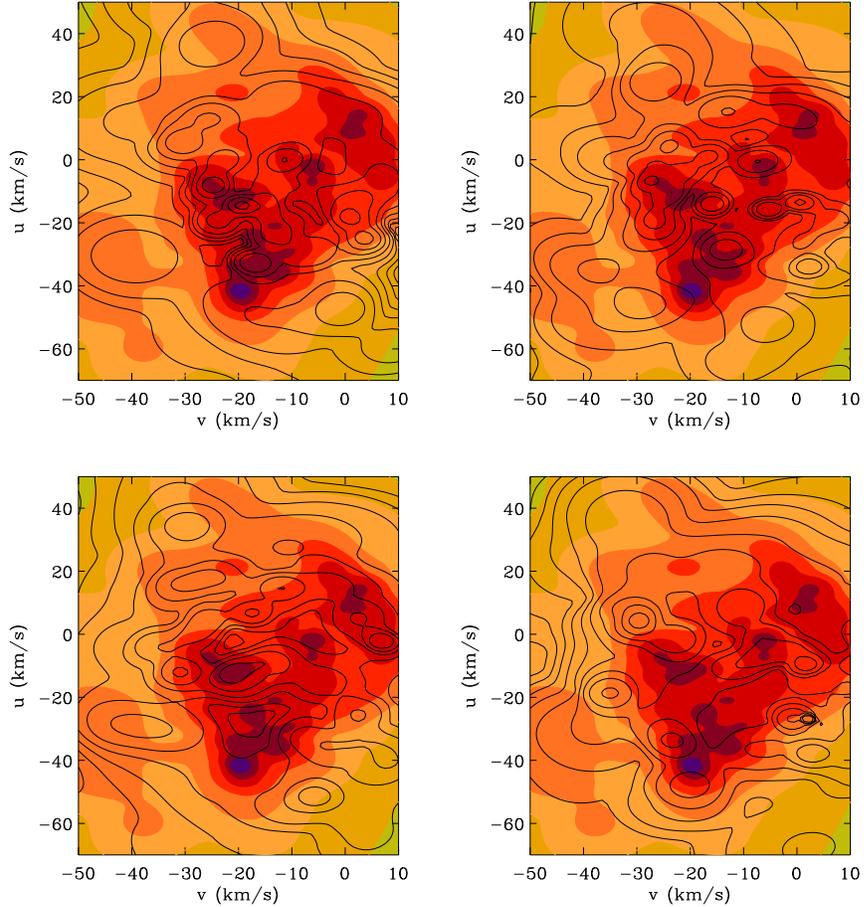}
\caption{\small{ Overlay of simulated velocity distributions, (when
the bar alone perturbs a warm exponential disk), on the observed local
velocity distribution, (shown by filled contours). The radial (in
units of $R_{\rm CR}$) and azimuthal coordinates of the observer
corresponding to the different panels starting from the top left, are
(1.8625, 25$^\circ$), (1.9025, 15$^\circ$), (1.9625, 35$^\circ$) and
(2.0875, 25$^\circ$). The distributions were smoothed using the
adaptive kernel method discussed in Skuljan, Hearnshaw $\&$ Cottrell
(1999). Logarithmic contours have been used with inwardly increasing
shading in the observed distribution. $U$ is positive along the
Galactic centre while $V$ is positive along the sense of Galactic
rotation. The ratio of the maximum field due to the bar at OLR to that
due to the background disk is about 3.6$\%$.}  }
\label{fig:test2}
\end{figure*}

\subsection{Goodness of Fit}
\label{sec:stats}
\noindent
The comparison between $f_o$ and $f_s$ could perhaps be carried out
visually, in terms of the degree of overlap between the $U-V$
coordinates of the 5 moving groups that are marked in
Figure~\ref{fig:fuxuv}, within the error bars given by the solar
peculiar velocities. However, this is hardly satisfactory; there is no
direct means of quantifying this ``degree of overlap''. Moreover,
given the radial and azimuthal ranges that we span in our runs, and
the widths of each $R$ and $\phi$ cell, each simulation generates
24$\times$9 model distributions in all the $R-\phi$ cells. Undertaking
a visual comparison of each of these 216 $U-V$ distributions to the
observed one, is firstly tedious and secondly subjective.

An alternative approach might be to carry out a test in order to
ascertain if $f_o$ and an $f_s$ are independent. The
Kolmogorov-Smirnov analysis is a typical example of such a statistical
test that can check for independence of distributions. However, our
distributions are bivariate $-$ this hinders the implementation of the
Kolmogorov-Smirnoff scheme since it is far from robust for
distributions more complicated than univariate ones. Additionally, our
distributions are highly non-linear in the sense that there are parts
of the $U-V$ space that are much more densely populated than
others. This renders the task of judging independence of the two
distributions even more troublesome.

Such logistical obstacles motivated us to adopt the less elegant
frequentists' view: we formulated a goodness of fit test that would
test the (null) hypothesis H(0) that: {\it the observed $U-V$ data is
drawn from the model distribution}. The conclusion is based on the
$p$-value of a test statistic.

The $p$-value of a statistic is such that if it is smaller than or equal
to the pre-set significance level of a test (usually taken as 0.05),
then the null hypothesis is rejected at that level. Here, the
significance level is the maximum probability that the statistic would
be as observed, assuming H(0) to be true.

We now discuss the method of estimating the $p$-value for a test
statistic {\cal{\bf S}} which is a decreasing measure of the goodness
of fit, i.e. the better the fit, smaller is {\cal{\bf S}}. The
reciprocal of the likelihood serves the purpose in this regard, as
suggested in \citet{prasenjitbook}. We define the likelihood of a data
set $D$, given a model distribution ($f_M$) at the physical location
$R-\phi$ , as:
\begin{equation}
{\bf prob}(D|f_M) = \prod\nu_{R,\phi}(U_i,V_j), 
\end{equation}
where $\nu_{R,\phi}(U_i,V_j)$ is the value of the velocity
distribution function at the location ($R$, $\phi$), in the
$(i,j)^{th}$ $U-V$ bin and the product is performed over all the $U$
and $V$ velocity bins.

Let $n$ (where $n$ is a natural number), data sets be drawn from a
model distribution. Let the test statistic defined on one such data
set be the reciprocal of the likelihood ({\cal{\bf S}}), such that for
the $i^{th}$ of the $n$ data sets it is {\cal{\bf S}}$_i$. For the
observed data set, let the statistic be ({\cal{\bf S}}$_o$). Then, we
try to compute the fraction of the $n$ data sets that fit less well
than the observed data, i.e. we monitor that for how many $i$, the
following inequality holds:
\begin{equation}
{\cal{\bf S}}_i \displaystyle{>} {\cal{\bf S}}_o
\end{equation}
If this is true for $m$ cases, then the $p$-value for the statistic
{\cal{\bf S}} is $m/n$. Thus, if 96$\%$ of the $n$ model data sets fit
better than the real data, then the $p$-value is 4$\%$; the model is
then rejected at 5$\%$ significance. In general, low $p$-values imply
that the null hypothesis is unlikely to be true. 

We identify a simulated distribution to be ``good'' if leads to the
maximum possible $p$-value (of 100$\%$); the $R-\phi$ location at
which this $f_s$ is recovered is then a ``good'' location.  A
distribution of $p$-values at each of the considered physical
locations is plotted on the $R-\phi$ plane; the estimate of the
$p$-value in any $R-\phi$ cell is shown as proportional to the darkness
of the shading used in that cell. The darkest cells are then used to
generate the $\pm$1-$\sigma$ ranges in $R$ and $\phi$ of
the ``good'' locations. (The distribution of these ``good''
locations is marginalised over $\phi$ to predict the range in solar
radius while the same, when marginalised over $R$, gives us the range
on the bar angle.)

We also check for the scatter in the distribution of the
``good'' locations in the $R-\phi$ plane, obtained for any model, to
confirm if the exercise undertaken here is a viable way of
constraining solar position.

Even though the $p$-value is subject to a number of criticisms that
can be taken into account by adopting Bayesian
techniques,\footnotemark~ it suffices to implement it in this case,
since we compare the $p$-values of discrete velocity distributions
that are binned into the same small number of velocity bins.

\footnotetext{ In the context of Astrophysics, the shortcomings of the
$p$-value and the superiority of the Bayesian inference are discussed
in an article by \citet{loredo}.}

\subsubsection{Comparison Between Models}
\noindent
Now it is to be stressed that this method of securing constrains on
the ``good'' locations, applies for a given model, i.e. a given
perturber added on a pre-fixed disc. This method also potentially
allows us to seek superiority of one model over another - if the
$p$-value at the location ($R_i$, $\phi_j$) in one model is less than
that in another, then we can say that the address ($R_i$, $\phi_j$) is
more akin to the solar position according to the former model than in the
latter. In this way, we can reject one model in preference to another,
only if the $p$-value in the former model exceeds that in the latter,
for all $i$ and $j$. Alternatively, if the distribution of $p$-values
over all $R$ and $\phi$ in one model is similar in shape to that in
another model, {\it and} the $p$-value from the former model at ($R_i$,
$\phi_j$), exceeds that from the latter for most $i$ and $j$, we
consider the former model better than the latter. (This is why we
reject Model~5 with respect to Model~4; see below).

When the differences in $p$-values alone cannot discern the viability
of a model, we resort to independent dynamical considerations, in
order to establish the same.  The models can also be pitted against
one another by checking if the right $\sigma_U$, $\sigma_V$ and $l_v$
have been recovered at the solar radius.



%
%

\subsection{Recording the Orbits}
\label{sec:record}
\noindent
In all our distribution diagrams, radial velocity ($U$) as observed
from the Sun is positive towards the Galactic centre while transverse
velocity ($V$) observed from the Sun is positive in the sense of
Galactic rotation.

Usually, the orbits are recorded in the frame that rotates with the
perturber. Now, the phase of the bar potential is not a function of
radius, unlike the potential of the spiral pattern. Thus, the frame
that the orbits are recorded in the spiral-only simulations will be
static with the spiral at a certain radius but not at any other
radius.

In the bar and spiral simulations the orbits have been recorded in the
rotating frame of the bar. In this frame the spiral pattern is not
static. Therefore we had to choose time points when the orbits could
be recorded. We chose to record the orbits at those time points, when
at the location of the corotation due to the bar, the potential of the
spiral pattern is maximised, in the frame rotating with the bar.

It needs to be emphasised that the $f_s$ that we recover are not
snapshots in time but are rather averages over time. This picture is
therefore concordant with $f_o$, which presents an average over ages.

\subsection{Effect of Pattern Speed of the Spiral}
\label{sec:omega_sp}
\noindent
We want to be able to understand the structures that develop in
velocity space in our bar and spiral models, as a function of the
spiral pattern speed. In order to accomplish this, we invoke the
important result that has been represented in Figure~14 in Paper~I:
stars are depleted from around the $ILR_s$ and pushed outwards to
higher radii.

When the ratio of pattern speeds of the spiral and the bar is 25/55,
$ILR_s$ is at $R$=1.42$R_{\rm CR}$, which is well inside the radial
range being studied, ($R$=1.7$R_{\rm CR}$ to 2.3$R_{\rm CR}$). Now
stars are driven away from $ILR_s$ to higher radii. Thus in this case,
even near the inner edge of this annulus ($R$=1.7$R_{\rm CR}$), there
are enough stars on prograde orbits, to contribute to the formation of
relevant structures at large (negative) values of $V$, particularly
the Hercules stream.

When the ILR is in the middle of the annulus under examination, (ratio
of pattern speeds is 18:55) stars are depleted from around the
resonance location ($R$=1.97$R_{\rm CR}$) and pushed outwards. It is
shown by our bar and spiral simulations that the average number of
stars just inside $ILR_s$ is less than in the initial equilibrium disk
by about $20\%$. Thus, in this case we can expect the $f_s$ to suggest
the observed bimodality only beyond this $ILR_s$. In accordance to
this, we do not expect satisfactory overlap between $f_o$ and $f_s$
inside $ILR_s$. That is indeed what we find in our distributions. The
smallest radius, at which an acceptable overlap occurs is $R/R_{\rm
CR}=1.9625$.

If the pattern speed of the spiral arms is chosen so that $ILR_s$
almost coincides with $OLR_b$, (i.e. at $R/R_{\rm CR}\approx1.7$) then
the spiral is responsible for pushing stars into the annulus under
investigation, from a radial location that is almost sitting at the
inner edge of this annulus. However, the number of stars entering this
annulus from lower radii is not as large as when the ILR lies inside
the considered radial band (fastest spiral).

\section{Results}
\label{sec:distr}
\noindent
The results from the different simulations are presented in this section.

\subsection{Bar Only: Model~1}
\label{sec:test2}
\noindent
In this section, we present the simulations performed by perturbing
the warm quasi-exponential disk with a bar, which contributes a
gravitational field that is at most $3.6\%$ of the field due to the
disk, at $OLR_b$. The $U-V$ distributions recovered at some locations
have been presented in Figure~\ref{fig:test2}. These are as viewed by
an observer at a radial location outside $OLR_b$ ($\ROLR/R_{\rm
CR}\approx 1.7$), and at low azimuthal separations from the major axis
of the bar; these are examples of some ``good'' model distributions
that we spotted from those generated at different locations, at the
end of the run.

One example of a model distribution that is not ``good'' is presented
in Figure~ref{fig:bad}; this $U-V$ distribution is one that would
appear to the observer who is along the bar major axis and a radius of
1.8325$R_{\rm CR}$. As is evident from this figure, the bimodality in
the observed local velocity distribution is not reproduced in this
simulated distribution.

\begin{figure}
\centering
\includegraphics[width=7cm]{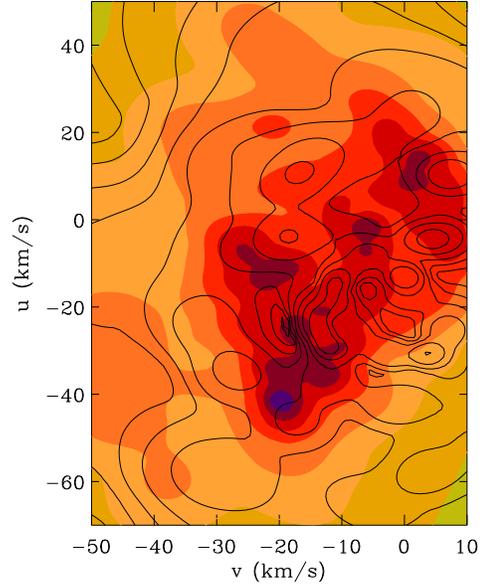}
\caption{\small{As in Figure~\ref{fig:test2}, except that in this
case, the observer location is (1.8325$R_{\rm CR}$, 0$^\circ$). At
this location, the Hercules stream is not well reproduced by the
modelled velocity distribution.}}
\label{fig:bad}
\end{figure}

\begin{figure}
\centering
\includegraphics[width=8cm]{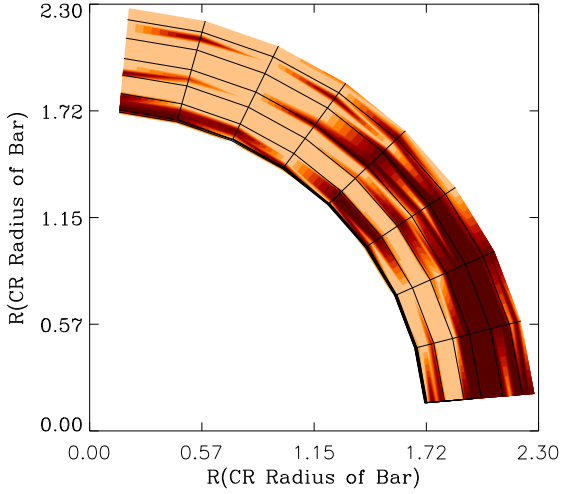}
\caption{\small{The distribution of $p$-value of the statistic
{\cal{\bf S}}, (see text), as recovered from the bar-only simulation
is represented as a contour plot over the relevant radial and
azimuthal ranges considered. The darkness in the shading increases
proportionally with $p$-value. Only the darkest $R-\phi$ cells (corresponding
to $p$-value=100$\%$) are used to constrain the solar position. }}
\label{fig:likely_bar6}
\end{figure}

The goodness of fit of the velocity distribution recovered in any
$R-\phi$ cell is represented by the $p$-value of {\cal{\bf S}} (the
reciprocal of the likelihood for the observed data to have been drawn
from the model distribution) in this cell. A contour plot of the
$p$-value of this statistic over the $R-\phi$ grid that we use, is
presented in Figure~\ref{fig:likely_bar6}.

From Figure~\ref{fig:likely_bar6}, we can see that in general, the
velocity distributions constructed at the lower azimuths comply with
the real $U-V$ data. In fact, by tracking the cells at which the fit
is the best, we can extract constraints on the bar parameters.

\begin{figure}
\centering
\includegraphics[width=7cm]{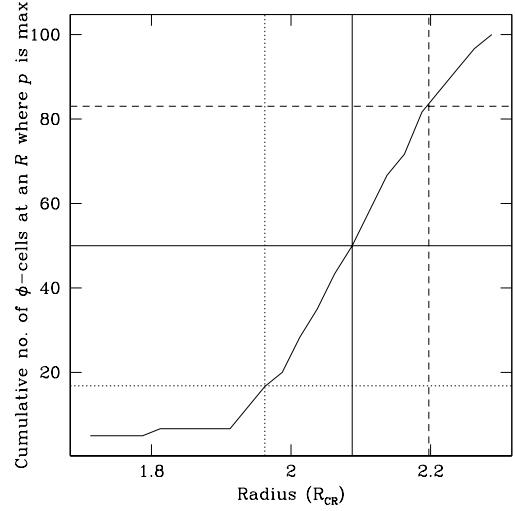}
\caption{\small{The cumulative total of the number of azimuthal bins
(expressed as a percentage of the total) at a given $R$, at which the
$p$-value of the statistic {\cal{\bf S}} is maximum, plotted as a
function of radius. The median is identified at 2.0875$R_{\rm CR}$
(solid lines), while $5/6^{th}$ ($\approx$1-$\sigma$) of the net total
is achieved at $R$=2.1975$R_{\rm CR}$ (dashed lines) and the
$1/6^{th}$ mark occurs at $R$= 1.9625$R_{\rm CR}$ (dotted lines). The
simulation in question is the bar only simulation.}}
\label{fig:cumu}
\end{figure}

Thus, we find that analysis of the spatial distribution of the
``good'' models suggests that the best radial location for reproducing
the observed velocity distribution is given by 2.0875$R_{\rm CR}$,
with $\pm$1-$\sigma$ errors spanning the range from 1.9625$R_{\rm CR}$
to 2.1975$R_{\rm CR}$. These estimates are obtained by calculating the
cumulative total of the number of azimuthal locations at a given
radius, for which the $p$-value has attained its maximum.
Figure~\ref{fig:cumu} represents the cumulative total of the obliging
number of $\phi$-locations, over the considered radial range.

\begin{figure}
\centering
\includegraphics[width=7cm]{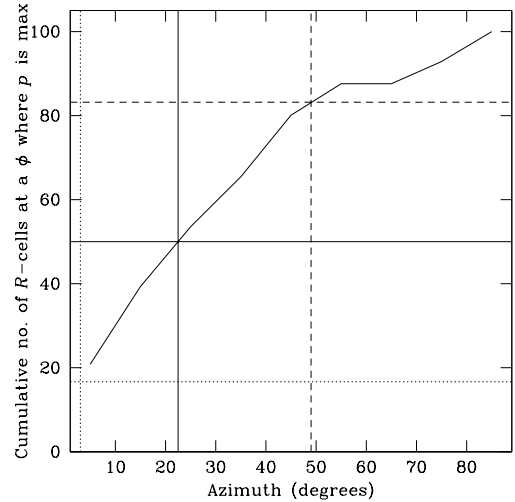}
\caption{\small{As in Figure~\ref{fig:cumu}, except that in this case,
the cumulative total of the number of radial locations (expressed as a
percentage of the total), at a given $\phi$, at which the $p$-value of
the statistic {\cal{\bf S}} is maximum, is plotted as a function of
azimuth. The median is identified at about 22$^\circ$, while
$5/6^{th}$ ($\approx$1-$\sigma$) of the net total is achieved at
$\phi$=49$^\circ$. All we can say about the $1/6^{th}$ mark is that it
occurs between 0 and 5$^\circ$.}}
\label{fig:cumu_th}
\end{figure}

When a similar construction is sought over the azimuthal range in
consideration, the result is shown in Figure~\ref{fig:cumu_th}. This
figure represents the run of the cumulative total of the number of
radial locations that bear the highest $p$-value, at a given $\phi$.
The best azimuthal location is noted to be about 22$^\circ$ while the
+1$\sigma$ mark occurs at about 49$^\circ$. It is hard to gauge the
azimuthal location corresponding to the -1$\sigma$ error since this
occurs inner to the obtained distribution.

\begin{figure}
\centering
\includegraphics[width=8cm]{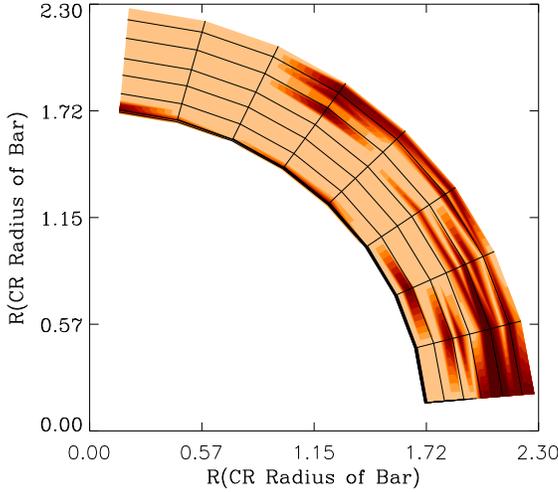}
\caption{\small{ As in Fig~\ref{fig:likely_bar6}, except that in this
case, a bar and a spiral pattern together disturb the disk. The ratio
of the pattern speed of the bar to that of the spiral is chosen to be
55:18 which places the $ILR_s$ at a radius of about 1.97 on a scale in
which the $OLR_b$ is at 1.7 approximately. }}
\label{fig:spiral18}
\end{figure}

The limits that the occurrence of the ``good'' models impose on the
radial and azimuthal locations of the observer, can be translated to
impose constraints on the bar parameters.

\begin{itemize}
\item The observer at the Sun needs to be separated from the
major-axis of the bar by an angle which lies in the 1-$\sigma$ range
of [0$^\circ$, 49$^\circ$], with the median of the distribution at
about 22$^\circ$.
\item The observer at the Sun needs to be constrained to the radial
range [1.9625, 2.1975]$R_{\rm CR}$ in our scale-free disk. If the
solar radius corresponds to the median of this distribution, then
$R_\odot = 2.0875^{+0.11}_{0.125}$. Here all radii are in units of
$R_{\rm CR}$. To relate these lengths to real units, let us consider
the solar radius to be $8.0$kpc. This scales $\ROLR$ to
$6.51^{+0.42}_{-0.32}$kpc and the corotation radius to
$3.83^{+0.25}_{-0.19}$kpc.
\end{itemize}
The constraint on the corotation radius can be translated to one on
the bar pattern speed. As discussed in Paper~I, the pattern speed of
the bar is set to unity in our analysis, (this is what renders the
corotation radius unity). Using the value of corotation radius deduced
above, and the value of $220$\kms$\:$ for the circular speed at the
Sun, the bar pattern speed is found to be about
$57.4^{+3.0}_{-2.5}$\kmskpc, in the Mestel potential of our
quasi-exponential disk.

\subsection{Bar and Spiral Pattern}
\label{sec:testspbar}
\noindent
A second set of simulations involved simultaneous stirring of the
warm, quasi-exponential disk by the bar and the spiral pattern.

The two different kinds of perturbers are responsible for the presence
of a relatively greater variety of families of stellar orbits in these
simulations. Intersection of the different generic orbits can cause
local density enhancements in the velocity space; this is manifest in
the numerous small clumps visible in the central parts of the velocity
distributions at many locations. However, the overall form of the
velocity distributions approaches the observed $U-V$ diagram; this is
also borne by the distribution of $p$-value of the statistic
{\cal{\bf S}} over the first quadrant of the outer part of the disc
that we study (Figure~\ref{fig:spiral21}).

\subsubsection{Lowest Ratio of Pattern Speeds: Model~2}
\label{sec:lowspiral}
\noindent
In Figure~\ref{fig:spiral18}, we present the distribution of 
$p$-value of {\cal{\bf S}, obtained by perturbing the warm
quasi-exponential disk by a bar and a spiral pattern
simultaneously. The maximum gravitational field due to the
perturbation is about $4.9\%$ of that due to the background
potential. The ratio of the pattern speeds is $\Omega_{\rm sp}:\Omega_{\rm bar}$=18:55.

From the distribution of $p$-value in this case
(Figure~\ref{fig:spiral18}), we can extract the locations where the
model velocity diagrams fit the observed one the best (i.e. where $p$
is highest). An analysis of such ``best'' locations is carried out
along the lines of the analysis discussed in the previous
section. This yields the following constraints from this simulation:
\begin{itemize}
\item The distribution of the highest $p$-value over radii, has its
median at $R$=2.0925$R_{\rm CR}$, with the +1-$\sigma$ and -1-$\sigma$
at $R$=2.21$R_{\rm CR}$ and $R$=1.95$R_{\rm CR}$,
respectively. Placing the Sun at the median of the distribution of the
best model locations with radius, we get the bar pattern speed to be
57.5$^{+3.2}_{-2.9}$\kmskpc.
\item The distribution of the highest $p$-value over azimuth, has its
median at $\phi$=15$^\circ$, with the +1-$\sigma$ and -1-$\sigma$
at $\phi$=30$^\circ$ and $\phi\in$[0,5]$^\circ$ respectively.
\end{itemize}
Thus, it appears that the ``best'' values of bar angles that this
simulation predicts is similar in range to that predicted by the
bar-only simulation. However, the radial range corresponding to the
``good'' models, is slightly skewed towards higher radii in this
case than in the other in the sense that higher radii are found
compatible with the solar position, in this case. This too is only to
be expected for the chosen pattern speed of the spiral which places
the ILR at about $R$=1.97$R_{\rm CR}$, thus causing depletion of stars
from around this radius, at the expense of enhancing the number of
stars at higher radii (see Section~\ref{sec:omega_sp}). 

\subsubsection{Highest Ratio of Pattern Speeds: Model~3}
\label{sec:highspiral}
\noindent
In these simulations, the ratio between the spiral and the bar
is maintained at 25/55. This places $ILR_s$ well inside $OLR_b$. The
ratio of the perturbative and background fields is about $0.049$ at
the $OLR_b$. The distribution of the $p$-value of {\cal{\bf S} is
shown in Figure~\ref{fig:spiral25high}. From
Figure~\ref{fig:spiral25high}, we learn that:
\begin{itemize}
\item the observer has to be constrained to radii between $R/R_{\rm
CR}$= 2.0875$^{0.15}_{0.25}$, in order to suggest satisfactory overlap
between the model and observed distributions. Here the errors are
1-$\sigma$ errors as usual. This implies that the bar pattern speed
is 57.4$^{4.1}_{6.9}$\kmskpc. 
\item The best value of the bar angle is 6$^\circ$, though this value
can vary within the 1-$\sigma$ range of [0$^\circ$, 43$^\circ$].
\end{itemize}
%

The bimodality observed in the local velocity distribution comes out
clearly in the distributions recovered from this simulation
(Figure~\ref{fig:barspiral_UV}). The other moving groups marked in the
local velocity distribution, are also reproduced.

\subsubsection{Intermediate Ratio of Pattern Speeds: Model~4}
\label{sec:midspiral}
\begin{figure}
\centering
\includegraphics[width=8cm]{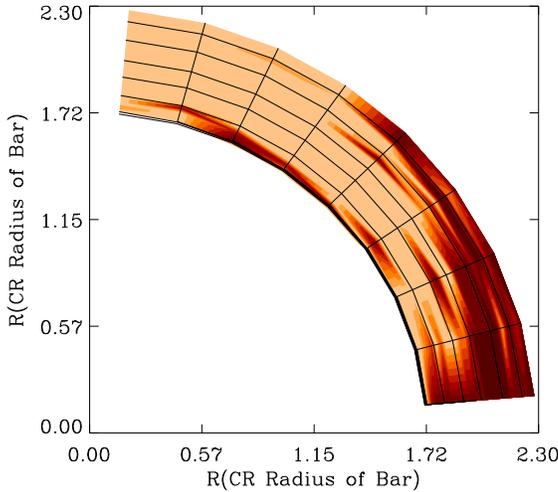}
\caption{\small{ Similar to Figure~\ref{fig:spiral18} except that the
pattern speeds of the spiral arms and the bar are in a ratio of $25:55$.}}
\label{fig:spiral25high}
\end{figure}

\noindent
The ratio of pattern speeds between the spiral pattern and bar is
chosen to be 21/55 in this simulation. This places $ILR_s$ almost on
top of $OLR_b$. The two perturbers together produce a gravitational
field that is at most $4.9\%$ of that due to the background
disk. Figure~\ref{fig:spiral21} shows the bivariate distribution of
$p$-value of the test statistic {\cal{\bf S}.

\begin{figure*}
\centering
\includegraphics[width=15cm]{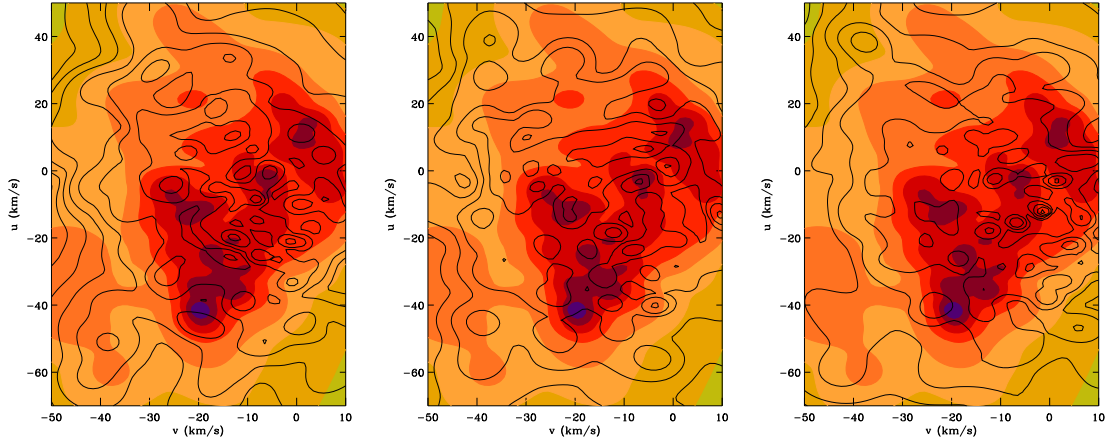}
\caption{\small{ Similar to Figure~\ref{fig:test2} except that in this
case, a spiral pattern disturbs the disc along with the bar; the
pattern speeds of the spiral arms and the bar are in a ratio of
$25:55$.  The observer locations for the above distributions are
(1.9625, 35$^\circ$), (2.0125, 25$^\circ$) and (2.1125, 25$^\circ$)
starting from the top left-hand panel.}}
\label{fig:barspiral_UV}
\end{figure*}

\begin{figure}
\centering
\includegraphics[width=8cm]{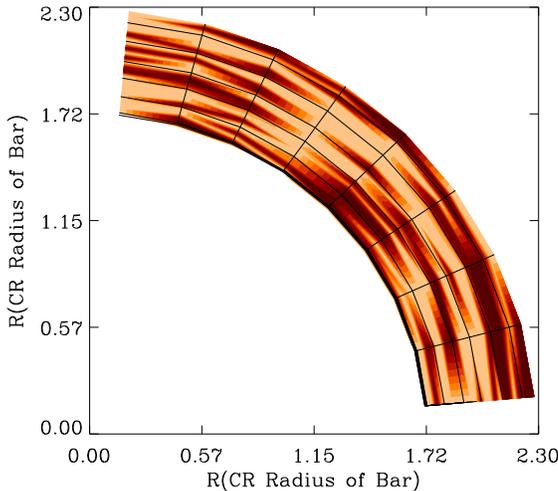}
\caption{\small{ Similar to Figure~\ref{fig:spiral18} except that in
this case, $ILR_s$ occurs almost at the same physical radius as
$OLR_b$. }}
\label{fig:spiral21}
\end{figure}
As is evident from Figure~\ref{fig:spiral21}, this distribution is
rather different from the distribution shown in
Figures~\ref{fig:spiral25high}, \ref{fig:spiral18} and
\ref{fig:likely_bar6}. The essential difference lies in the fact that
in this case, the regions corresponding to the highest $p$-value
are strewn all over the range of the $R-\phi$ space that we record the
orbits in. In the previous runs, the distribution of the locations
corresponding to the best models was much less non-linear. In fact,
this $p$-value distribution bears a strong resemblance to that which
results from a spiral-only simulation, (see below).

Since the ``good'' locations are too scattered in the $R-\phi$ plane,

it is, rather meaningless to try and use this simulation to extract
the relevant bar parameters.

\subsection{Spiral Only: Model~5}
\label{sec:spiral}
\noindent
Model velocity distributions were also obtained from simulations in
which the 4-armed spiral pattern alone perturbed the warm
quasi-exponential disk. The results discussed below pertain to
spiral-only simulations that were performed with all of the three used
pattern speeds.

The distributions obtained in this simulation, are characterised by
structures that stand out less boldly than those observed when the bar
is a perturber though the number of velocity space structures is
much more in this case. Now, the potential of the logarithmic spiral
has a radial dependence. At a given azimuth, stars on varying radii
are at different relative phases compared to the maxima in the
potential well of the spiral pattern. Thus, the cumulative effect at
this azimuth is in general due to the superposition of orbits with
different orientations. (At a given azimuth and at two close radii,
orbits will in general differ only slightly in orientation unless the
radii under consideration straddle a principal resonance
location). Such superposition of orbits can lead to local density
enhancements in velocity space. This would render the velocity
distributions inundated with structures on small length-scales in the
$U-V$ plane. 

In order to check if this formed structure is compatible with the same
in the observed distributions, we check the distributions of our
goodness-of-fit statistic (Figures~\ref{fig:only_spiral}). 
\begin{figure}
\centering
\includegraphics[width=8cm]{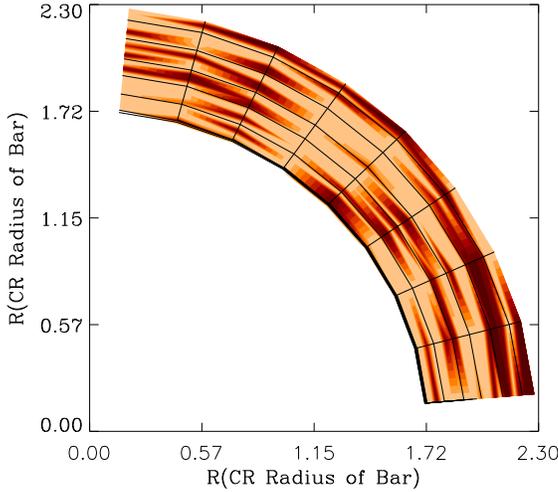}
\caption{\small{ Similar to Figure~\ref{fig:spiral18} except that in
this case, a spiral pattern perturbs the disk alone. The pattern speed
of the perturber is 18/55 times that of the bar and the used spiral
contributes a gravitational field that is at most 3.6$\%$ of that due
to the background disk, at $OLR_b$.}}
\label{fig:only_spiral}
\end{figure}

From the distribution of the $p$-value in this simulation, we notice
the following:
\begin{itemize}
\item There is a dearth of continuous bands of ``good'' locations;
instead, the $R-\phi$ cells corresponding to the ``good'' models, are
strewn all over the band under consideration. This is unlike the
results obtained from simulations done with the bar alone and the bar
and spiral simulations in which the slowest and fastest spirals have
been used (Figures~\ref{fig:spiral25high}, \ref{fig:spiral18} and
\ref{fig:likely_bar6}). The result presented in
Figure~\ref{fig:only_spiral} bears similarity to that obtained from
the simulation done with the bar and the spiral pattern of
intermediate speed (Figures~\ref{fig:spiral21}). An attempt will be
made in Section~\ref{sec:scatter}, to understand the origin of
this clumpy nature of the distribution.
\item The spiral alone is found to cause insufficient disk
heating. Using a higher perturbation strength does not improve the
situation much. This will be talked about in Section~\ref{sec:discussions}.
\item The highest $p$-value is lower in this case compared to when
the bar is included.
\end{itemize}
Given the highly non-linear form of the distribution of the maximum
$p$-value in this simulation, we will not resort to constraining the
solar position from this simulation. (The recovery of the bar angle
from this spiral-only run is of course not valid).

\subsection{Dispersions}
\noindent
It is important to check that the velocity dispersions of the
configuration at the end of a run, at a radius that can be identified
with the solar position, is compatible with that estimated in the
solar neighbourhood from observations. Needless to say, dispersions
vary with the choice of the stellar samples. However, the sample of
initial conditions that we numerically integrate is not distinguished
by differences in age or metalicity. Also, the $f_o$ that we compare
our $f_s$, is constructed from $all$ the single stars that obey
certain cutoffs in the Hipparcos Catalogue. Thus, our recovered values
of dispersions will be averages over all ages. All we hope for is that
the runs result in dispersions that lie within the ranges pertinent to
an average sample in the solar neighbourhood: from main sequence stars
to giants.  This is indeed found to be the case, as apparent from a
comparison of the values of $\sigma_U$, $\sigma_V$ and $l_v$ shown in
Figure~\ref{fig:disp} from the different models, with Table~10.2 and
Table~10.3 of \citet{BM} (pages~632-633). It is to be noted that the
presented quantities are azimuth averaged, at the medial value
estimated for the radius of the Sun, from the corresponding run. When
it is not possible to extract the solar position from a simulation,
the radius is chosen to be equal to that obtained from the bar-only
simulation.

It may be noted that while the vertex deviation at the solar radius is
acceptable for all the models, for the spiral-only run, the Oort ratio
($\sigma_V^2$:$\sigma_U^2$) is almost as low as the lower limit for
the old stellar disk (of about 0.42) from {\it Hipparcos}
\citep{jameswalter}. This is most probably an offshoot of the weakness
of the spirals that we work with, as explained in
Section~\ref{sec:method}.
\begin{figure}
\hskip-.4cm
\includegraphics[width=8cm]{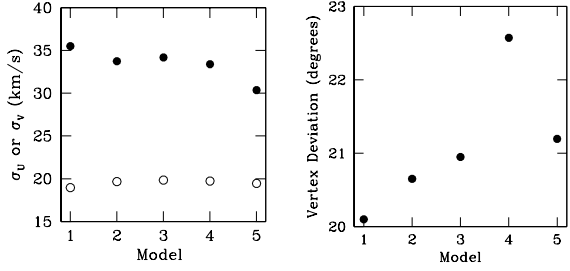}
\vskip-4.2cm
\caption{\small{Left panel shows plots of azimuth averaged radial
(filled circles) and tangential velocity dispersions (open circles),
as obtained at the recovered solar radius, at the end of the
simulation, for each of the five different runs that were
conducted. These are as follows: (1) bar-only (2) bar and fastest of
the three spiral patterns (3) bar and slowest of the spirals (4) bar
and spiral when the ILR of the spiral concurs with the physical
location of the OLR of the bar (5) spiral-only. Models 4 and 5 do not
allow a solar position to be extracted; in these cases, the
dispersions were extracted at an $R$ that was identified as the solar
radius in the bar-only simulation. The azimuth averaged vertex
deviation at the solar radius is shown in the right panel.}}
\label{fig:disp}
\end{figure}

\begin{table*}
\caption{Table showing the results obtained from the five different
runs that have been presented in this paper. The third, fourth and
fifth columns respectively depict the value of the azimuth averaged
$\sigma_U$ and $\sigma_V$ and $l_v$, at the end of the run, at a
radial location that is identified as the median of the 1-$\sigma$
range of radii that are found compatible with the solar radius. The
next two columns refer to the value of the bar pattern speed and
angle, as inferred from the relevant run. Models 4 and 5 imply highly
scattered distributions of the locations where the observed velocity
diagram is well reproduced (see text). These models therefore fail to
provide adequate constraints on the solar position and hence cannot be
used to extract the bar parameters.}  \hrule \centerline {
\begin{tabular}{c|c|ccccc}
Number & Model & $\sigma_u$ (kms$^{-1}$)& $\sigma_v$ (kms$^{-1}$) & $l_v$ (degrees)& $\Omega_{\rm bar}$ (kms$^{-1}$kpc$^{-1}$)& Bar Angle (degrees)\\
1 & bar-only & 35.51 & 18.96 & 20.1 & 57.4$^{+3.0}_{-2.5}$ & 22$^{+27}_{-22}$\\
2 & bar $\&$ spiral, $\Omega_{\rm sp}$ = 25/55$\Omega_{\rm bar}$ & 35.75 & 19.67 & 20.6 & 57.4$^{+4.1}_{-6.9}$ & 6$^{+37}_{-6}$\\
3 & bar $\&$ spiral, $\Omega_{\rm sp}$ = 18/55$\Omega_{\rm bar}$ & 34.18 & 19.84 & 21.0 & 57.5$^{+3.2}_{-2.9}$ & 15$^{+15}_{-15}$\\
4 & only spiral, $\Omega_{\rm sp}$ = 25/55$\Omega_{\rm bar}$ & 33.40 & 19.74 & 22.6 & $--$ & $--$ \\
5 & bar $\&$ spiral, $\Omega_{\rm sp}$ = 21/55$\Omega_{\rm bar}$ & 30.37 & 19.45 & 21.2 & $--$ & $--$ \\
\end{tabular}
}
\hrule
\vskip.3cm
\label{tab:results}
\end{table*}

\section{Orbits}
\label{sec:orbits}
\noindent
It is expected that just outside the $OLR_b$, stars are on
orbits which are aligned with the major-axis of the bar while just
inside, anti-aligned orbits prevail. The radial location of the Sun,
it is also expected to be visited by stars which are scattered off the
Outer $1:1$ Resonance, (the $-1:1$ resonance). This resonance is
defined by the following rule.
\begin{equation}
-1 = \frac{\kappa}{\Omega - \Omega_{\rm p}}
\label{eq:-1:1}
\end{equation}
Here $\kappa$ and $\Omega$ are the epicyclic and azimuthal frequencies
while $\Omega_{\rm p}$ is the pattern speed of the perturbation.
Eqn.~\ref{eq:-1:1} implies that this resonance occurs at a radius of
about $2.41$ in units of the corotation radius. It is possible that
stars can reach the solar circle from the vicinity of this
resonance. Thus, near the Sun, we can expect orbits belonging to the
the x$_1$(1) family, (orbits aligned to the bar), the x$_1$(2) family
at negative $U$ and $V$ velocities, (anti-aligned orbits) and the
$-1:1$ type orbits. We also expect a plethora of chaotic orbits,
especially around the location of the OLR. The nomenclature used here
is that of \citet{contopop}. 

\subsection{Stars on ``hot'' orbits}
\noindent
The orbit in the left panel of Figure~\ref{fig:orb1} is due to a star
at position coordinates $R/R_{\rm CR} =2.04$ and azimuth 35$^\circ$),
at very large negative radial and tangential velocities, ($U=-81$\kms,
$V=-40$\kms). The velocities characterising this orbit are high enough
to place this star well outside the Hercules stream, (towards more
negative radial velocities), in the $U-V$ plane. This orbit can be
ruled out as chaotic on the basis of the smooth surface of section
(right panel in Figure~\ref{fig:orb1}. This orbit is noted to librate
around a closed aligned orbit.

\begin{figure*}
\centering
\includegraphics[width=12cm]{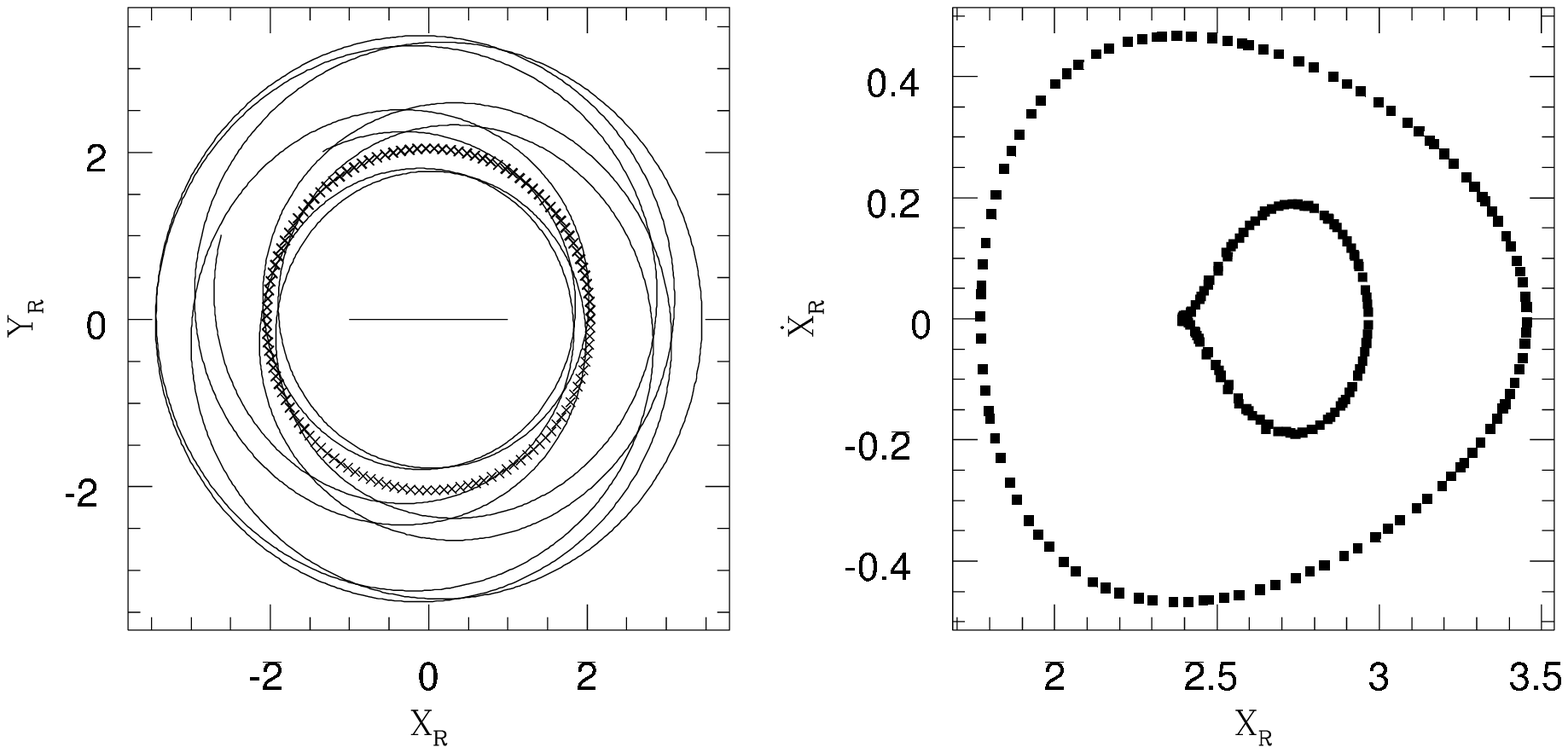}
\caption{Figure showing orbit (left panel) and its surface of section (right 
panel), of a star with initial velocities $U=-81$\kms, $V=-40$\kms,
 near the solar radius. This orbit is parented by a closed aligned
 orbit. The circular orbit at this radial location is marked by
 crosses. The bar is static in this frame that rotates with it.
 The major axis of the bar is shown by the solid
 horizontal line. This orbit corresponds to an energy in the rotating
 frame of the bar, that just exceeds the range of energies
 that characterise orbits that are susceptible to cross the corotation
 and escape to much larger radii. We observed such orbits only at
 radial velocities less than -70\kms$\:$ in the $U-V$ plane; such a
 constraint on $U$ suggests that only regions outside the Hercules
 stream, (as marked in Figure~\ref{fig:fuxuv}) are due to these
 ``hot'' orbits. The bar is strong enough to contribute a
 gravitational field at the OLR that is at most $3.6\%$ of the field due
 to the background potential. }
\label{fig:orb1}
\end{figure*}
The parent of this orbit could have originated just outside the OLR.
It is also possible in principle, that the birthplace of the closed
aligned orbit was inside corotation; at a radius just inside
corotation, stars are sired primarily by closed aligned orbits. A star
can foray to the solar radius from such radii, only if it is highly
energetic. The orbit presented in Figure~\ref{fig:orb1} is indeed
characterised by a very high energy. Thus it is possible that this is
one of the ``hot'' orbits which \citet{raboud} and \citet{fux} claim
to be the building blocks of the Hercules stream.
To ascertain the birth place of the star which is on the orbit in
Figure~\ref{fig:orb1}, we decided to examine its energy. In the
rotating frame of the bar, the Jacobi integral ($J$) provides the
value of the Hamiltonian.
\begin{equation}
J = \frac{U^2}{2} + \frac{(V + v_0 -\vert{{\bf R}\times{\bf \Omega_{\rm p}}}\vert)^2}{2} + \Phi_{\rm eff}
\label{eqn:jacobi}
\end{equation}
Here $U$ and $V$ are velocities recorded, with respect to the Sun,
$v_0$ is the amplitude of the uniform rotation curve and $\Phi_{\rm
eff}$ is the effective potential, given by:
\begin{equation}
\Phi_{\rm eff} = \epsilon_0\frac{\cos(2\phi)}{R^3} + \ln{(R})
-\frac{\Omega_{\rm p}^2}{2}R^2  
\label{eqn:poteffec}
\end{equation}
where, $\epsilon_0$ is the maximum strength of the bar ($=0.06$) and
$\phi$ is the azimuth. It may be noted that in Eqn~\ref{eqn:jacobi},
the solar peculiar velocities have been ignored. If the value of the
Jacobi integral exceeds the effective potential at the unstable
Lagrange points, a star inside corotation is in principle able to
escape to infinity. It is the Coriolis term in the equation of motion
(written in the rotating frame), that in general prevents the star
from doing so. But even so, such stars can easily visit the solar
neighbourhood, from a location inside corotation. \citet{fux} and
\citet{raboud} call such stars to be on ``hot'' orbits.

To check if the orbit shown in Figure~\ref{fig:orb1} is ``hot'', we
first identified the Lagrange points for the effective potential
defined in Equation~\ref{eqn:poteffec}. The value of the Hamiltonian
for a star on this orbit was then calculated and compared against the
value of $\Phi_{\rm eff}$ at the saddle points of this potential
structure. We found that this orbit is just energetic enough to be
termed ``hot''; the Jacobi integral characterising this orbit exceeds
the effective potential at the saddle points (which are also the
unstable Lagrange points) by about $0.123\times{10^3}{\rm km}^2{\rm
s}^{-2}$.

A neighbouring orbit ($U$=-81/kms and $V$=-43/kms) is definitely
hot. It is shown in Figure~\ref{fig:orb2}. The surface of section of
this orbit manifests its chaotic nature. The Jacobi integral exceeds
the effective potential at the saddle points in $\Phi_{\rm eff}$ by
$1.08\times{10^3}{\rm km}^2{\rm s}^{-2}$.
\begin{figure*}
\centering
\includegraphics[width=12cm]{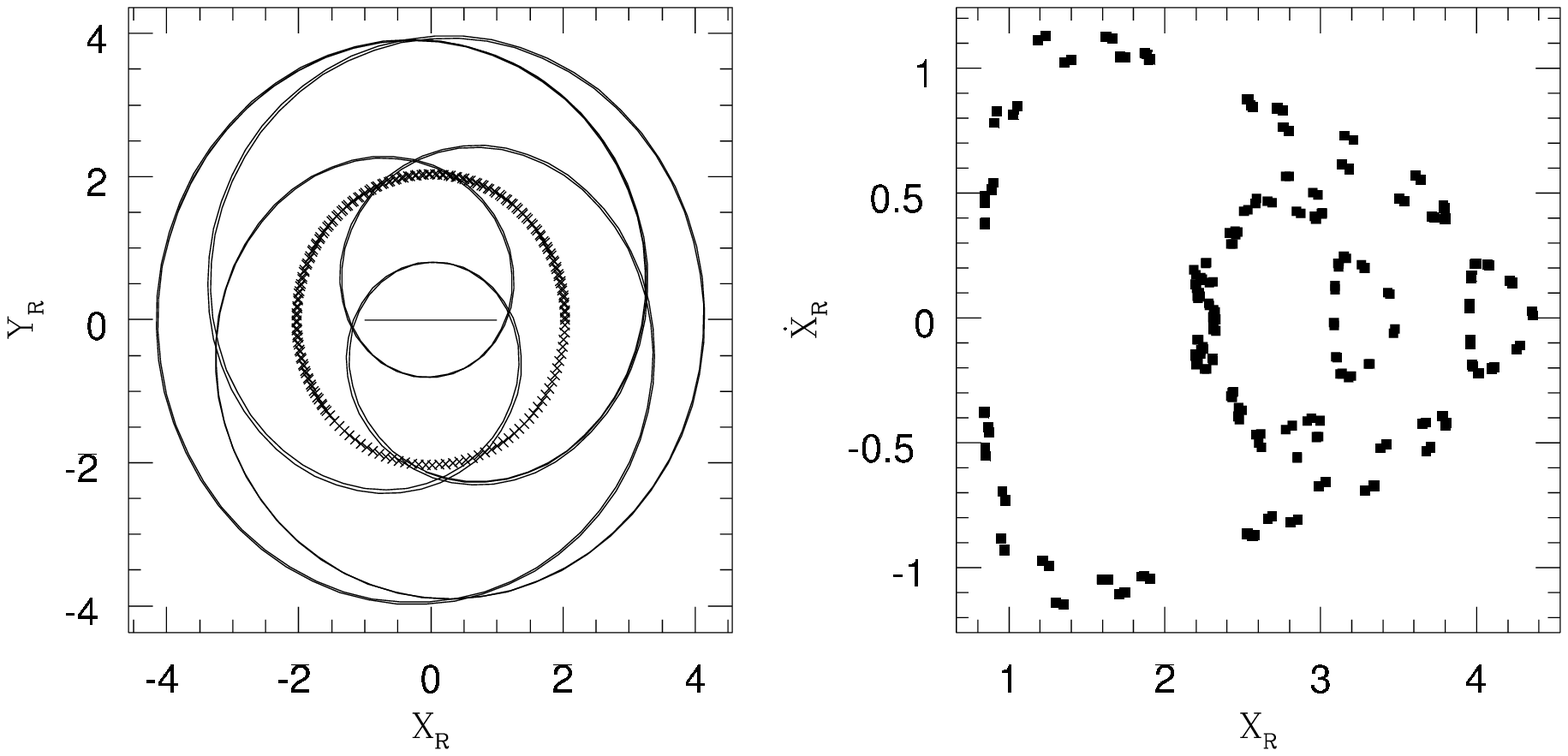}
\caption{Similar to Figure~\ref{fig:orb1}. This is the orbit charted
 out by a star with velocities $U=-81$\kms, $V=-43$\kms, near the
 solar radius. The orbit is definitely irregular as implied by its
 surface of section; no smooth curve was found to join the occupied
 islands of phase space. Considerations of the Hamiltonian of this
 orbit suggested that this is a ``hot'' orbit.}
\label{fig:orb2}
\end{figure*}
Thus, our conclusion is that the progenitor of the orbit in
Figures~\ref{fig:orb1} and \ref{fig:orb2} lie inside corotation radius
in the initial equilibrium disk and stars on these orbits are pushed
to much larger radial locations, owing to their energies in the frame
of the bar. 

{\it However the ``hot'' orbits cannot be considered responsible
for the Hercules stream since they lie substantially outside the
Hercules stream on the $U-V$ plane}. On the other hand, such a
conclusion is subject to the definition of the exact boundaries of the
Hercules stream. It is rather difficult to search for exact
confinement of individual moving groups in the local velocity
space. According to the local velocity distribution diagram presented
in \citet{walterolr}, the star on this ``hot'' orbit would indeed lie
within the region of velocity space that is dominated by late-type
stars. Dehnen refers to this mode of the local velocity distribution
as the ``OLR'' mode; his OLR mode is the analogue of Fux's Hercules
stream. Differences between the methods used by Dehnen and Fux in the
construction of the velocity distribution function and the considered
stellar samples, were responsible for variation in the details of the
respective distribution diagrams. Thus Dehnen's OLR mode covers
greater area in the $U-V$ plane than Fux's Hercules stream. 

In light of the above discussion, it appears that the following will
be safe to conclude: {\it we spotted stars at very high radial
velocities which were found to be on ``hot'' orbits, and did not find
any stars on such orbits at $U\leq70$\kms}. The moderate to highly
``hot'' orbits were found to be chaotic.

\subsection{Hercules stream}

\noindent
A number of stars in the Hercules group were observed to be on
quasi-periodic orbits belonging to the anti-aligned family though it
appears that this stream is composed of chaotic orbits too.  We found
a plethora of chaotic orbits in this group - the closer the observer
is to the principal resonance due to the bar, the greater is the
fraction of such chaotic orbits.

\subsection{Hyades-Pleiades stream}
\noindent
It appears that the less energetic stars in this group are due to the
outer $1:1$ resonance. In addition to the $-1:1$ family, there are
also orbits of the aligned family.  By constructing surfaces of
sections for stars in this part of the $U-V$ plane, we have realised
that in addition to these orbits, there are also a number of highly
irregular orbits that constitute the Hyades-Pleiades group.

Of course, once the spiral arms are introduced along with the bar, the
families mentioned above interact with the orbits generic to the
region around $ILR_s$. As is borne by the distributions presented in
Section~\ref{sec:testspbar}, this affects the velocity space by
introducing local density enhancements and washing out prominent
clumps. The structure of phase space in this case is better understood
in light of a brief orbital analysis when the spiral wave is the sole
perturber. This is dealt with in the following section.

\section{Discussion}
\label{sec:discussions}
\noindent
This section is devoted to detailed discussion of the results
presented above.
\subsection{Origin of the Scatter in the $p$-Value Distribution}
\label{sec:scatter}
\noindent
As we have seen above, the spatial distribution of the $p$-value of
{\cal{\bf S} is marked by a high degree of scatter, for the
spiral-only simulations and the run in which the major resonances of
the two perturbers coincide. For the other simulations, the locations
corresponding to the ``good'' models are less spread out over the
quadrant of the radial band that we study. In this section, we
question the origin of this behaviour. 

It is envisaged that such a trend can be addressed only in terms of
the orbits that characterise the model velocity distributions. We
investigate the orbits in different $R-\phi$ cells, for the
spiral-only simulation in details. This simulation is preferred over
the special bar and spiral simulation mentioned above, since it is
relatively easy to interpret the results when there is a single
perturbation instead of two. Besides, given that we are dealing with a
4-armed, tightly wound spiral pattern, the effect is nearly
axisymmetric; therefore, we expect the azimuthal dependence of the
velocity distributions (at a given $R$), to be relatively less in the
spiral-only case.

We note the following trends in our results:
\begin{itemize}
\item Trend~1-\\
A direct anti-correlation is found to exist between
$p$-value of the statistic and the evenness in the distribution of the
loop orbits between the apocentric and pericentric radii. Thus, when
the orbit fills out the annulus between these two radii almost
uniformly (left panel in Figure~\ref{fig:orbit_sp}), {\it $p$-value is
lower than when the orbits falls on itself consistently}, i.e. the same
path is repeated in configuration space (right panel in
Figure~\ref{fig:orbit_sp}).
\item Trend~2-\\
At any radius, the azimuth-averaged $p$-value of {\cal{\bf S}} is
found to roughly vary periodically with $R$. This is shown in
Figure~\ref{fig:p_R}.
\item Trend~3-\\
The overall distribution of the $p$-value is not affected by the
pattern speed of the spiral used in the spiral-only simulation; the
strength of the clumps in this distribution may vary slightly with the
position of $ILR_s$.
\item Trend~4-\\
Beyond $R\approx$1.8$R_{\rm CR}$, the distribution of the
``good'' models in the spiral-only case look very much like that in
the bar and spiral simulation, when the two resonances concur. In
addition to the approximate radial periodicity in the occurrence of the
``good'' models, the distributions of the $p$-value are noted to be
highly scattered in both Models~4 and 5.
\end{itemize}

\begin{figure*}
\centering
\includegraphics[width=12cm]{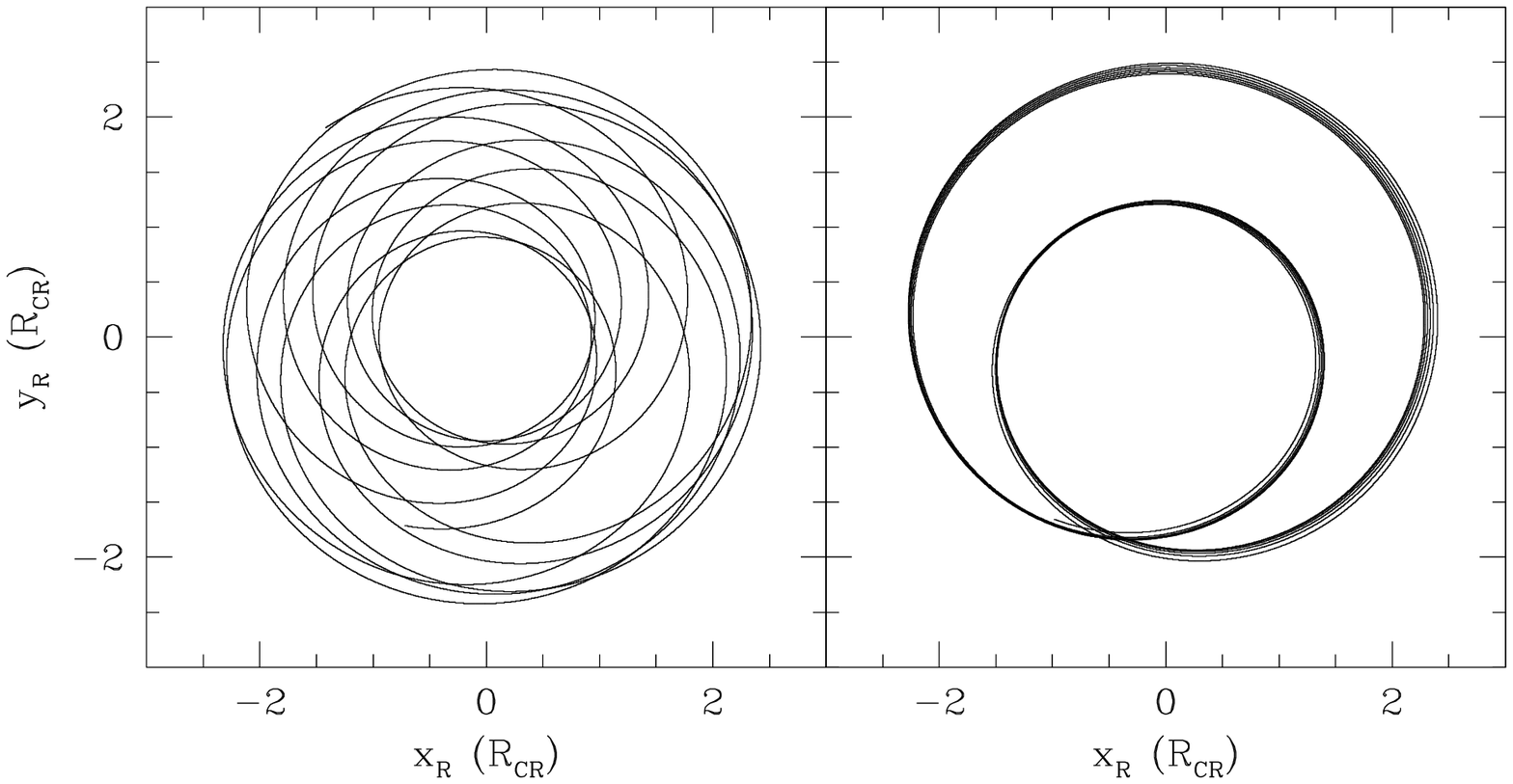}
\caption{\small{Orbits in the presence of the perturbation due to the
spiral alone, at an azimuth of 35$\circ$ and radii of 1.8625 (left panel)
and 1.9375 (right), on a scale in which the ILR of this spiral pattern
occurs at a radius of 1.42 approximately. Both are stable orbits.}}
\label{fig:orbit_sp}
\end{figure*}

\begin{figure}
\centering
\includegraphics[width=7cm]{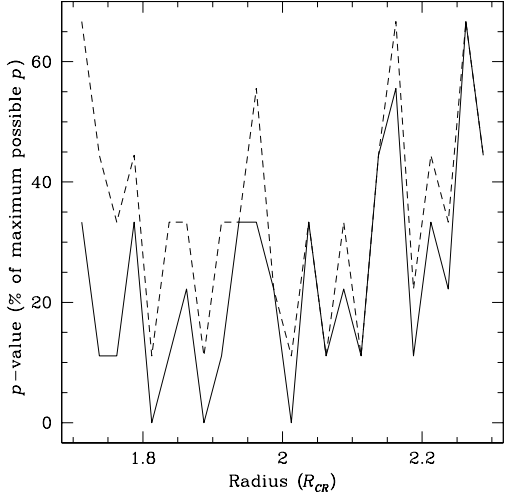}
\caption{\small{Azimuth averaged $p$-value of {\cal {\bf S}}, plotted
as a function of radius for the case of the spiral-only simulation
(solid lines) and the resonance overlap case (broken lines). In both
cases, the variation in the $p$-value is noted to be similar except
that its amplitude is higher in the case when the resonances
coincide.}}
\label{fig:p_R}
\end{figure}

Now we attempt to qualitatively understand the 4 trends itemised in
the last paragraph.

\begin{itemize}
\item Understanding Trend~1-\\
It is quite straightforward to realise why the even-ness of
distribution of the orbit within the pericentre and apocentre should
be strongly anti-correlated with the $p$-value. When the orbit is as
in the left panel in Figure~\ref{fig:orbit_sp}, the resultant of the
velocity vectors, taken over varying phases, implies a radial velocity
that is clustered strongly around zero and $V$ clumped preferentially
at a high positive value, if the sense of motion along the orbit is
positive, otherwise the peak of $V$ is strongly negative. In general,
this picture will not correspond to the definite structures that are
observed at distinct locations on the $U-V$ plane, in the solar
neighbourhood.

\item Understanding Trend~2-\\ However, the main question is why the orbit folds
upon itself at certain radii but deviates from this configuration at
other radii. In fact, the variation of $p$-value with radii, in a
spiral-only simulation is found to be roughly periodic, as shown in
Figure~\ref{fig:p_R}. We need to understand the origin of this
periodicity.

Whether an orbit is going to fold upon itself or it is going to be
highly excursive will be given by the initial conditions and some kind
of an integral of motion, which must in turn relate to the
potential. In our scale free units, the perturbing potential of the
spiral is:
\begin{equation} 
\Psi_{sp}(R, \phi) \sim \displaystyle{\frac{\cos[4(\phi - \Omega_{\rm sp}\Delta{t}) -
                                                 \alpha\ln{R}
                                                ]
                                           }{\sqrt{R}}
                                     }
\end{equation}
where $\Omega_{\rm sp}$ is the pattern speed of our 4-armed spiral pattern,
$\Delta{t}$ is the time that has elapsed and $\alpha$ is related to
the pitch angle $i$ of the spiral as $\alpha=4\:{\cot}(i)$.  This implies
that whatever the exact form of the relevant integral of motion is, it
will be roughly periodic in $\ln{R}$.

Thus, the volume between the pericentre and the apocentre is given by
an integral of motion that is akin to the component of the angular
momentum that is orthogonal to the plane of the disc ($L_z$). In the
axisymmetric case, it is indeed given by $L_z$, where
\begin{equation}
L_z^2 = \displaystyle{\frac{1}{R^3}
                      \left(\frac{\partial\Psi_{sp}}{\partial{R}}\right)
                     }
\end{equation}
Thus, it is clear that in our spiral only simulation, the radial
extent of excursion that these orbits are allowed, varies periodically
with the initial radial position of the star on this orbit. This is
indeed what is noted from the run. It is to be noted that the
discussion above included only stable orbits.

\item Understanding Trend~3-\\ This explanation would hold good irrespective of
$\Omega_{\rm sp}$ used.

\item Understanding Trend~4-\\ Regular orbits are harder to spot in the case of the
spiral and bar jointly perturbing the disc, with their major
resonances coinciding. In this configuration the distribution of
$p$-values resembles that of the spiral-only case and it {\it does
not} look like that in the other three models. The variation of the
azimuth averaged $p$-value at any radius varies with radius nearly
periodically, (plot in broken lines in Figure~\ref{fig:p_R}) while
this quantity is not at all periodic with $R$ in the other cases. Why
is this so?

Qualitatively, we may try to understand this trend by querying what it
is between the spiral-only and the resonance overlap cases that is
common, which is but different from the other three models?  Well, in
these two cases, the dynamical influence of the spiral pattern is not
swamped by that of the bar, unlike the other three cases. Now, the
coincidence of the major resonances at $R\approx1.7R_{\rm CR}$ is an
example of {\it resonance overlap}. As \citet{fordwalker} pointed out,
when multiple resonances occur sufficiently closely, the system is not
only rendered nonintegrable, but extensive chaos also sets in. This is
what is predicted to happen in this case, as suggested previously by
\citet{quillen}. So is it possible that global chaos is triggered
in the spiral-only case too but the introduction of the bar inhibits
this effect? To answer this question, we need to recall what the
spiral does, as distinguished from what the bar does.

As we have mentioned before, stars are drifted outwards from
$ILR_s$. The bar on the other hand, distorts orbital configurations
with respect to its major axis. Thus, if we have an $R-\phi$ cell that
is being visited by orbits of opposite orientations, then it is
possible that a systematic decrement occurs in the value of one of the
components of the velocity vector. In the spiral-only case though, the
orbits that superpose are nearly of the same configuration; such
superposition leads to an enhancement in the velocity. Thus, the
kinetic energy keeps building up in the spiral only case, and in the
process of this build-up, if the kinetic energy crosses a threshold,
then it is possible that chaos sets in $-$ chaos that is strong enough
to disrupt the orbital families that support the spiral. Once such
chaos sets in, the superposition of the velocity vectors of the stars
at different phases on it may coincidentally add up to produce a
feature in velocity space that is akin to one of the observed
streams. There is no reason to expect the ``good'' locations to be
confined to well defined bands when this happens.  This is a
qualitative argument to explain the high scatter that is noted in our
spiral-only runs.
\end{itemize}


\subsection{Distributions}
\noindent
In the $U-V$ distributions that are identified as ``good', it is found
that the simulations replicate the observed velocity space structures
quite well. More importantly, the locations that correspond to {\it
low (lower than maximum) $p$-values} for the considered statistic,
bear {\it distributions that do not visually resemble the observed
local velocity distribution}. This offers confidence in the usage of
the $p$-value of {\cal{\bf S}} as an indicator of the goodness of fit.

In the ``good'' distributions, the bimodality is found to be clearly
marked. In such $f_s$, as in $f_o$, the Hercules stream represents
 a smaller probability density in the velocity space than
the other four streams. At lower radii, greater structure is observed
than at higher radii; in particular the region of the velocity space
around the moving groups, Hyades, Pleiades and Coma Berenices is
heavily crowded with structure.

Other than the 5 moving groups that we focus upon in this work
(namely, Hercules, Hyades, Pleiades, Coma Berenecius, Sirius), in most
of the ``good'' distributions, we consistently notice two structures
which are not currently distinguished as moving groups:
\begin{itemize}
\item A clump often shows up in the ``good'' distributions, at about
$U=20$\kms and $V=-20$\kms. 
\item All the good model distributions, display a finger-like feature
that extends from about $U=40$\kms and $V=0$\kms to about $U=50$\kms
and $V=-20$\kms. This feature is also noted in the observed local
$U-V$ diagram. Perhaps, this could be explained as a low-density
moving group, such as the Hercules stream.
\end{itemize}

\subsection{Relative Significance of Bar and Spiral}
\label{sec:significance}
\noindent
It is found that the bar alone can reproduce the observed velocity
structures quite well, though the bar alone simulations (visually)
appear to be responsible for clumps that are bolder than those spotted
in the $U-V$ plane in the bar and spiral simulations, (compare
Figure~\ref{fig:test2} to Figure~\ref{fig:barspiral_UV}). In fact, the
extra wispiness which characterises the distributions from the
bar+spiral simulations, seems to render the $U-V$ diagrams more close
to the observed distribution. However, such a judgement is based on
visual comparison and is therefore purely qualitative since our
goodness of fit parameter is not able to deal with this facet of
the comparison. Thus, a visual check may suggest that the bar only
simulations are not sufficient to model local kinematics, but unless
the goodness of fit parameter is refined to include appreciation of
the sub-structure of the clumps in velocity space, it cannot confirm
any such conclusion. At the moment, the comparison is made merely on
the basis of the position and extent of these clumps, within error
bars given by the solar peculiar velocity components.

Inclusion of the spiral arms reduces the boldness of the features that
show up in $f_s$. We realised that the presence of the extent of fine
structure in the distributions was a function of the chosen ratio of
pattern speeds of the spiral wave and the bar. With the appropriate
choice of $\Omega_{\rm sp}$, we can ensure that more stars enter the solar
neighbourhood from lower radii when the models included the spiral
pattern than when it was left out. We could use this ratio as a tool
to control the prominence of the Hercules stream. Smoothening of the
local phase space by even other scattering agents could further
improve the overlap between the model local distributions and their
observed counterpart.

Another reason why the spiral pattern is efficient in diluting the
boldness of the structures is actually due to the way we choose to
record the orbits. The orbits are recorded in the rotating frame of
the bar, in which, the spiral pattern is obviously not static. Neither
is the phase of the potential of the logarithmic spiral the same, with
respect to that of the bar, at all radii; the phases of the two
potentials coincide only at the corotation due to the bar. Thus, in
general, within a given radial bin, superposition of orbits of
slightly different phases occurs. This leads to a slight washing off
of the bold features in the velocity plane, causing an increase in the
wispiness of the velocity structures.

However, we can convince ourselves of the importance of including the
spiral pattern in the modelling in another way. The interarm
separation of the spiral pattern used in our work is about 3.4kpc,
given that the pitch angle of this pattern is $15^\circ$, the number
of arms is 4 and that the Sagittarius arm lies at about 6.5kpc
\citep{melnik, vallee02}. If the average epicyclic excursion
of stars in the Solar neighbourhood exceeds this interarm separation,
then spiral arm scattering can be ruled out as important. Now, if the
epicyclic amplitude of a star with a guiding centre at $R_{\rm G}$ is
$X$, then after time $t$, the star is at a radius
\begin{equation}
R = R_{\rm G} + X\cos{(\kappa{t})}
\label{eqn:epicyclic}
\end{equation}
Here $\kappa$ is the epicyclic frequency which is given by $\ds{\frac{
\sqrt{2}}{R}}$ in the Mestel potential. 
We can set the average $\dot{R}$ to the radial velocity dispersion
$\sigma_{\rm U}$, which is about 35\kms at the solar circle, after the
perturbation has steadied. Thus, taking the time derivative and then
the averages of both sides of Eqn~\ref{eqn:epicyclic}, we get that
$X\approx\ds{\frac{\sqrt{2}\sigma_{\rm
U}}{\kappa}}\approx1.25$kpc. Therefore, on the average, the total epicyclic
excursion of a star in our simulations is about 2.5kpc. This is less
than the inter-arm separation of the spiral pattern used in our work.
Thus the scattering induced by the spiral pattern in our models cannot
be ruled out. Hence any modelling that does not take such effect into
account is incomplete.

\subsection{Origin of the Moving Groups}
\noindent
The extensive set of stellar kinematical information that {\it
Hipparcos} provided, has opened up the possibility of constructing
velocity distribution diagrams in our neighbourhood. While workers
differ in the details of this portrayal \citep{walterdf, fux, skuljan,
chereul}, one feature that is clearly evident is a bimodality in the
local velocity distribution; this can also be identified as the
location of the Hercules stream in the $U-V$ plane, away from the
other moving groups. As \citet{famaey05} indicate, there are three
broad classes of mechanisms that have been invoked in this context,
with the aim of describing the observed moving groups, and in
particular, the emergence of this bimodality.

The most traditional of these is to correlate the moving groups with
irregularities in the star formation rate. However, such a scenario
fails to reconcile with the spread in age shown by stars within a
moving group. Discussions about the span of ages present in any of
the streams is beyond the scope of our current methodology. The
inclusion of transient spirals or episodic growths in bar strength in
the modelling, can help in this regard.

Another future project is envisioned in which the radial velocity and
metalicity information of a very large number of nearby stars, as
provided in the RAdial Velocity Experiment (RAVE) data set
\citep{steinmetz_rave}, supplemented by transverse velocity data of
the same stars from the Hipparcos measurements, will be implemented
into an algorithm like the one we used here, in order to identify the
original galactocentric locations of these stars that results in a
metalicity/age distribution that is observed today. As the quality of
kinematic data in the solar neighbourhood improves, we will have
better constraints to check our models against.

A merger scenario, could also be invoked to explain the moving groups
observed in the disk in our neighbourhood; the merger in question is
between the Milky Way and a satellite galaxy. After all, this
mechanism has been solicited to explain streams in the halo, near the
solar position \citep{helmi} and the Arcturus group
\citep{navarro04}. However, given that these distinct moving groups are
observed in the disk, it would be (statistically speaking) improbable
for them to be all merger remnants.

Alternatively, it is the dynamical influence of the bar that is
sometimes called upon to explain the moving groups. As explained
above, Kalnajs attributes the bimodality in the local $U-V$ diagram to
orbits that have been scattered off the OLR of the bar, as recorded by
an observer at the Sun, sitting near or just outside this
OLR. \citet{walterolr} agrees with this while \citet{fux} suggests
that the Hercules stream is made up of ``hot'' orbits that have been
discussed in Section~\ref{sec:orbits}. \citet{fux_aa} proposes that
the Hercules stream is due to an overdensity of chaotic orbits; the
chaos resulting from the bar achieving a major resonance in the
vicinity of the solar radius. A plethora of chaotic orbits have been
spotted in our bar-only simulations too.

\subsubsection{Why the Hercules stream?}
\noindent
In the light of our results, it appears that though the Kalnajs
mechanism can adequately explain the very visible bimodality in the
local velocity space, it is not relevant to the splitting of the
structure at the less negative tangential velocities (i.e. the
velocity mode other than the Hercules stream) into the four (or more)
individual moving groups. Strictly speaking, even the origin of the
Hercules stream is not as clean as the Kalnajs mechanism suggests
since in our bar-only simulations, we find that there is greater
variety besides the anti-aligned orbits. At the same time we cannot
agree that there are only ``hot'' orbits in the Hercules stream; in
fact, the ``hot'' orbits manifest themselves only at $U\leq-70$\kms.
Thus, the hypothesis proposed by \citet{raboud}, \citet{fux} and
\citet{fux_aa} is not satisfactory either.

The bar used in our simulations was imposed on a warm
quasi-exponential disk. An important observation that can help us to
resolve the puzzle of the origin of the streams is that {\it our
simulations done with a much weaker bar on a cold Mestel disk produced
a very clean and well-defined bimodality in the velocity space around
the location of the OLR of the bar}; one part of the $U-V$ plane is
occupied by aligned orbits while the other part is due to anti-aligned
orbits. For this setup, it was noted that there are very few stars
that were not members of these groups of aligned and anti-aligned
orbits. Also, {\it these two groups in the velocity plane appeared
coherent and not split into further sub-clumps}. All this is clear
from the distributions presented in Paper~I.

These findings lead us to the following conclusions:
\begin{itemize}
\item The presence of the aligned and anti-aligned modes in the
distributions obtained from the cold background model suggests that
the Kalnajs mechanism is certainly effective in the formation of the
Hercules stream.
\item In contrast to the observed velocity distribution, the lack of
splitting of either of the aligned or anti-aligned groups in the cold
disk scenario suggests that {\it as the underlying distribution of stars in
the disk becomes hotter and the bar develops to higher strengths,
stars will tend to get more energetic and will be less likely to be
trapped in a family of a certain orbital configuration}. Instead, stars
in the radial regions close to the Sun will then be more susceptible
to the more minor resonances, such as the -1:1 resonance that occurs
outside the solar radius. This will be manifest in the high degree of
structure noticeable in the $U-V$ plane.
\end{itemize}

Thus, we propose that a good way to view the situation in the solar
neighbourhood is to consider it as a progression from the case of the
cold disk being perturbed by a weak bar towards a configuration marked
by higher perturbation strengths and increased background stellar
velocity dispersions.

We should stress the fact that the cold disk simulations in question
had a $1/R$ surface density profile. As the observer moved away from
the OLR to higher radii, the prominence of the anti-aligned family
decreased in the velocity plane. However, such an observer location
will continue to be visited by stars from the inner disk, more so if
the surface density profile is changed so that it is approximately
exponential. Then, even at the solar circle, (which is outside the OLR
of the bar) there will be more stars coming from lower radii than in a
disk with a $1/R$ profile; these will contribute strongly to the
Hercules stream. Thus the Kalnajs mechanism is the {\it basis of the
bimodality of the velocity distributions around a principal resonance
of an $m=2$ perturbation}. In the presence of parameters realistic to
the solar neighbourhood, other effects show up.

\subsubsection{What of Model~1?}
From the distributions that we recover from our bar-only simulations,
it seems that the bar alone is sufficient to explain the moving
groups. However, as we have seen above
(Section~\ref{sec:significance}), the average epicyclic excursion of a
star (locally) is less than the inter-arm separation of the four-armed
spiral pattern typical of our Galaxy. This implies that the modelling
of local kinematics is incomplete without the effects of the outer
spiral pattern being taken into account. In other words, even though
our goodness of fit parameter indicates otherwise, dynamical
reasoning suggests that Model~1 is not satisfactory.

\subsubsection{What of Model~5?}
The spiral only runs lead to a value of $\sigma_U\approx30$\kms,
$\sigma_V/\sigma_U\approx0.66$ and $l_v\approx21^{\circ}$ in the disc,
at the solar radius. This combination of values is neither compatible
with the counterparts of these quantities for any of the $B-V$ colour
bins into which \citep{jameswalter} divide the {\it Hipparcos} sample
of main sequence stars nor with the same for the non-main sequence
stars the kinematic information of which is given in Table~10.3 in
\citet{BM}. However, it is always possible that these quantities that
we recover at the solar radius are a result of averaging over certain
ages or colour bins; that such an averaging over age results in higher
$\sigma_U$ from all the other 4 models, compared to Model~5, indicates
that Model~5 fails to produce enough disc heating. Thus, the dynamical
viability of this model is questionable. Even if the dispersions from
the spiral only models compared favourably with the observations,
these models would still be inapplicable in any attempt to constrain
the solar position, owing to the scattered nature of the ``good''
locations.

While this may be considered to be a direct result of the weakness of
the spiral pattern that defines this model, it also needs to be
stressed that Paper~I indicates that spiral strength has to be much
higher, (more than double of the currently used strength) to achieve
an acceptably large $\sigma_U$.

Moreover, the $p$-value distribution for Model~5 and 4 are very
similar; when we compare the amplitude of these distributions, we find
that the maximal $p$-value attained in the spiral only simulations,
falls short of 100$\%$ in most of the $R-\phi$ bins. Hence we can
expect that the spiral only simulations do not reproduce the observed
phase space structure, at least as efficiently as Model~4. 

\subsubsection{What of Models~2, 3 and 4?}
\noindent
Other than this paper and Paper~I, it is the work of \citet{quillen}
that reports the results of simulations that account for both the bar
and a spiral wave. \citet{quillen} identifies the inclusion of the
spiral pattern with a large fraction of chaotic orbits, when the solar
radius corresponds to a location near or just outside $ILR_s$. She
suggests that this scenario corresponds to a spiral pattern speed of
0.75 times the angular rotation rate at the Sun, i.e. $\Omega_{\rm sp}
\approx 0.75v_0/R_0 \approx 21$kms$^{-1}$kpc$^{-1}$. This value
matches the pattern speed, used in our work, that places $OLR_b$ at
the physical location of $ILR_s$. But in contrary to the suggestion
that resonance overlap causes the Hercules stream, we notice {\it
``good'' $U-V$ distributions even for slower as well as faster
spirals} (Figures~\ref{fig:spiral18} and \ref{fig:spiral25high}.  In
other words, Models~2 and 3 are indeed successful in producing all the
5 stellar streams that are observed in the Milky Way disc, in the
solar neighbourhood.

However, when the resonances of the two perturbations coincide, the
locations where the modelled $U-V$ diagram matches the observed one,
are found to be scattered all over the disk, rather than being
confined to a well-behaved band, (Figure~\ref{fig:spiral21}) unlike in
the runs done with the other $\Omega_{\rm sp}$. This invalidates the usage
of this technique of extracting values of bar characteristics from
such a model. Thus, even though Model~4 is found to reproduce the
observed structures in velocity space, it cannot be used to constrain
the solar position.

\subsubsection{Which $\Omega_{\rm sp}$ should we expect?}
\noindent
An important question that emanates from this conclusion is the
possibility of the coincidence of the locations of the major
resonances of the bar and spiral pattern in the disk of our Galaxy.
Is there a relevant dynamical mechanism that would motivate such a
scenario, i.e. couple the bar and the outer spiral pattern in this
way? Thus, for example, a spiral pattern that is joined to the ends of
the bar (the inner spiral pattern, which we do not deal with here), is
expected to share the pattern speed of the bar. We do not know of any
dynamical phenomenon that would lock the bar and the outer spiral in
such a configuration, but would advance investigation into this area
as potentially interesting. The point is, that if the pattern speeds
of the two structures do indeed relate in this way, then we would not
in principle be allowed to extract solar position from comparing
velocity distributions at different $R-\phi$ addresses.

\subsection{Transient Spirals}
\noindent
The spiral pattern in our work was grown adiabatically to its maximum
strength over a time that matched the growth time of the bar. This is
most probably not correct and a transient spiral pattern would have
been more realistic. Since our perturbations were grown adiabatically
to their maximum strengths, the resulting velocity distributions are
not time dependent. The effects of a transient spiral arm was reported
by \citet{fux_aa}, \citet{quillen} and \citet{tremainewu}; it is not
surprising that in the presence of such a feature, the $U-V$
distributions vary with time.

In fact, \citet{tremainewu} carry out their direct integration of test
particles in a sheared sheet, as perturbed by stochastic spiral waves
alone. They suggest this non-axisymmetric perturbation to be the
source of the moving groups and more importantly, the ability of the
spiral pattern to aid in the radial migration of stars (see Paper~1,
discussion above, \citep{binneysellwood} is invoked to explain the
range of ages noticed within a moving group.  A crucial advantage of
including a spiral wave in the modelling is that the ability of the
spiral pattern to push stars across radii implies that the presence of
a wide range of ages within the same moving group can be explained. Of
course, such modelling is incomplete without the bar simply because
the central bar in our Galaxy exists, and its effects will be felt
strongly around its OLR.

Thus, it appears that a satisfactory dynamical explanation for the
origin of the 5 stellar streams in the solar neighbourhood is
definitely available as due to the joint handiwork of both the bar and
the spiral pattern.

\subsection{Criticism of Backward Integration}
\noindent
As an aside, it merits mention that the simulations carried on in this
paper were direct in time, as distinguished from the formalism
advanced by \citet{walterolr} - this is an integration scheme that is
conducted backwards in time, starting from the current scenario in the
solar neighbourhood. Inspite of its obvious benefits, we do not
advocate this procedure, based on the results of experiments in which
the perturbation was grown in a disk, maintained at the saturation
strength for some time and subsequently dissolved (see Paper~I for
greater details). It was found that no matter how slowly the growth
and dissolution were carried out, the resulting velocity distributions
were consistently marked with ``holes'' in them. This observation has
been discussed in Paper~I and stems from the loss of those orbits that
come too close to the resonance, (becoming chaotic thereby). This
effect was first noted by \citet{binneyspergel}. Given the prominence
of dynamical resonances in our simulations, it is expected that
backward integration will lead to a spurious velocity distribution.
Hence we adopt direct integration instead. \citet{fux_aa} suggests
that the backward integration technique picks out fine-grained
distribution function when it is really the coarse-grained one that
should be compared to distributions. Additionally, \citet{famaey05}
mention that since the streams observed among the $K$ and $M$-giants
are young kinematic features, backward integration is an invalid mode
of treatment.


\section{Conclusions}
\noindent
We have presented results of simulations in which a warm
quasi-exponential disk has been perturbed by a bar or an outer spiral
pattern or by both these structures acting simultaneously. The aim of
these simulations is to model local kinematics; in particular, we aim
to reproduce the five moving groups - Hercules, Hyades, Pleiades, Coma
Berenecius and Sirius. Additionally, we aim to verify the validity of
the very attempt to constrain the solar position from a comparison of
the simulated and observed velocity distributions.

For a given model, this comparison is quantified in terms of a
goodness of fit measure: $p$-value of the statistic \cal{\bf S}},
which is the reciprocal of the likelihood that the observed data was
drawn from the simulated distribution at hand. The arrangement of the
$p$-values on the $R-\phi$ plane is monitored to extract the
distribution of the $R-\phi$ addresses with the highest value of
$p$. Wherever $p$-value is maximum, $f_s$ is closest to $f_o$. We
calculate the median of this bivariate distribution, as well as the
$\pm$1-$\sigma$ boundaries, first at a given radius and then at a
given azimuth. This enables us to place constraints on where the
observer at the Sun is, with respect to the major axis of the bar, in
order for the observed local kinematics to be reproduced best. We can
then translate these constraints to bar parameters, namely, the bar
angle and the bar pattern speed.  

The main results of this paper are enumerated as follows.
\begin{enumerate}
\item 
The spiral only model failed to produce good enough fits and enough
disc heating, prompting us to reject Model~5; the dynamical influence
of the central bar is important even at the solar radius and cannot be
ignored in kinematic modelling of the solar neighbourhood. It may be
argued that the dearth of disc heating that we find is a result of our
choice of weak spirals; experience from Paper~I suggests that even
doubling the spiral strength would not have helped, though even
stronger, transient spirals have been found to work
\citep{tremainewu}.
\item The bar only model corresponds to multiple ``good'' velocity
distributions that is reflected well in the $p$-value distribution,
but considerations of the interarm separation of the 4-armed Milky Way
spiral indicates that this exceeds the average radial excursion of a
star in the bar only simulations. In other words, proper kinematic
modelling requires the inclusion of the spiral - so we reject Model~1.
\item For the first time, local kinemtical modelling has been
undertaken while scanning through a range of this ratio; the observed
moving groups were reproduced even when $ILR_s$ is well separted from
$OLR_b$, as well as for resonance overlap.  In other words, all models
that include the spiral pattern along with the bar, work.
\item In Table~\ref{tab:results}, the results of the simulations are
enumerated, along with the azimuth averaged dispersions and vertex
deviations in the disk, at the solar radius, at end of the run. Within
the $1-\sigma$ errors, the overlap of the ranges indicated by the
different, usable runs, suggest a bar pattern speed of
57.4$^{+3.0}_{-2.5}$\kmskpc and a bar angle that lies in the range
[0$^\circ$, 30$^\circ$]. These values sit comfortably with what is
suggested by \citet{debattista} and \citet{peter}, though (as with the
$\Omega_{\rm bar}$ from these works) our recovered pattern speed is in
excess of what is suggested by \citet{walterolr}.
\item This work indicates that even though the endeavour of
identifying the solar position from kinematical modelling has been
reported before in the bar only case \citep{walterolr}, blind
implementation of such exercise is found to be no longer viable. We
find that this exercise to be viable only in the cases when the
dynamical influence of the spiral is outweighed by that of the bar,
i.e. in the bar only case or the case of bar+spiral simulations, in
which {\it resonance overlap does not occur}, i.e. as long as
$\Omega_{\rm sp}$ does not lie in the immediate vicinity (inside
$\pm2.5\%$) of 21/55 times $\Omega_{\rm bar}$. Thus, as long as
$\Omega_{\rm sp}$ is not about 22\kmskpc, all bar+spiral simulations
can be used to constrain the solar position.
\item Although attempts have been made to constrain the solar position
(and thereby the bar parameters) from kinematical modelling with the
bar alone, \citep{walterolr, fux}, the comparison of the simulated and
observed velocity distribution has not been hitherto quantified. This
renders the investigation of multiple models hugely laborious and such
comparison essentially subjective. The statistical means of
identifying the ``good'' models, as advanced in this paper, is an
objective alternative.
\item A high degree of scatter is noticed in the distribution of the
``good'' locations on the $R-\phi$ plane, for the spiral only and
resonance overlap cases.  It is predicted that the same will happen
with strong spirals, in the bar and spiral case, even when the
resonances of the bar and spiral are seperated. This behaviour is
conjectured to be due to the tendency of the spiral to push stars away
from its ILR, as compared to a bar, which distorts orbital orientations
leading to a reduction in the net kinetic energy of an ensemble of
test particles (i.e stars). Such examination of the effects of the
spiral, as distinguished from that of the bar, is presented here for
the first time.
\item We realise that bimodality in the local $U-V$ plane can be
induced even in models in which a weak bar perturbs an initially cold
disk. In this case, the bimodality is caused by the Kalnajs mechanism,
as is evident from the presence of only anti-aligned orbits in the
family at more negative transverse velocities and only aligned orbits
in the other, at a radial location very close to $OLR_b$. The further
the observer moves from $OLR_b$, the weaker the anti-aligned group
becomes.
\item As the perturbation strength and the warmth of the background
disk increases, we would expect the fraction of chaotic orbits to
increase; the two main groups are then noted to split into smaller
sub-clumps. Thus, the splitting of the group at the less negative
$V$ values can be understood to happen due to this enhancement in the
chaotic orbits.
\item With an increase in dispersion, as the stars get more energetic,
they become more susceptible to other minor resonances such as the
outer -1:1 resonance (Equation~ref{eqn:-1:1}). Interaction of orbits
from such varying resonant families also encourage the splitting.
\item Analysis of orbits from the bar only simulations show that stars
at $|U|\leq$70\kms~ in the Hercules stream are found to be not
``hot''.
\end{enumerate}

\begin{acknowledgements}
\noindent
The author is supported by a Royal Society Dorothy Hodgkin
Fellowship. I acknowledge the help of Prof. Michael Merrifield whose
suggestion to establish a goodness-of-fit parameter, in preference to
relations between bivariate distributions, made this work possible. I
also wish to thank Dr. Wyn Evans and Prof. James Binney for their
suggestions and criticisms. The author is indebted to Dr. Roger Fux
for supplying the data of the local velocity distribution and the
parameters used in the smoothing algorithm that is used to extract the
distribution from this data.\\
\end{acknowledgements}

\vspace{1cm}
\noindent

\bibliographystyle{aa}

\begin{thebibliography}{}

\bibitem[Block \& Puerari(1999)]{blockpuerari} 
Block, D.~L., \& Puerari, I.\ 1999, \aap, 342, 627 

\bibitem[Binney $\&$ Merrifield(1998)]{BM}
Binney, J. and Merrifield, M., 1998, {\em Galactic Astronomy}, Princeton University Press Princeton New Jersey.

\bibitem[Binney et. al(1997)]{binney97}
Binney, J., Gerhard, O. E. and Spergel, D. N., 1997, \mnras, 288, 365.

\bibitem[Binney $\&$ Tremaine(1987)]{bible}
Binney, J. and Tremaine, S., 1987, {\em Galactic Dynamics}, Princeton University Press Princeton New Jersey.

\bibitem[Binney $\&$ Spergel(1984)]{binneyspergel}
Binney, J. and Spergel, D., 1984, \mnras, 206, 159.

\bibitem[Bissantz et. al(2003)]{bissantz}
Bissantz, N. and Englmaier, P. and Gerhard, O., 2003, \mnras, 340, 949.

\bibitem[Boutloukos $\&$ Lamers(2003)]{botoloukas}
Boutloukos, S. and Lamers, H., 2003, \mnras, 338, 717.

\bibitem[Chakrabarty(2004)]{paper1}
Chakrabarty D., 2004, \mnras, 352, 427.

\bibitem[Chereul $\&$ Grenon(2001)]{chereul01} 
Chereul, E. and Grenon, M., 2001, {\em Dynamics of Star Clusters and the Milky Way}, eds. Deiters, S., Fuchs, B., Just, A., Spurzem, R. and Wielen, R., ASP Conference Series Vol 228, 398.

\bibitem[Chereul et. al(1998)]{chereul}
Chereul, E., Crézé, M. and Bienaymé, O., 1998, \aa, 340, 384.

\bibitem[Contopoulos $\&$ Grosbol(1989)]{contopop}
Contopoulos, G. and Grosbol, P., 1989, \araa, 1, 261.

\bibitem[Debattista et al.(2002)]{debattista} 
Debattista, V.~P., Gerhard, O., \& Sevenster, M.~N.\ 2002, \mnras, 334, 355.

\bibitem[Dehnen(1998)]{walterdf}
Dehnen, W., 1998, \aj, 115, 2384. 

\bibitem[Dehnen(1999)]{walter99}
Dehnen, W., 1999, \apj, 524L, 35. 

\bibitem[Dehnen $\&$ Binney(1998)]{jameswalter}
Dehnen, W. and Binney, J., 1998, \mnras, 298, 387. 

\bibitem[Dehnen(2000)]{walterolr}
Dehnen, W., 2000, \aj, 119, 800. 

\bibitem[de Simone et. al(2004)]{tremainewu}
De Simone, R., Wu, X. and Tremaine, S., 2004, \mnras, 350, 627.

\bibitem[Eggen(2004)]{olina}
Eggen, J. O., 1996, \aj, 111, 1615.


\bibitem[Englmaier \& Gerhard(1999)]{peter} 
Englmaier, P., \& Gerhard, O.\ 1999, \mnras, 304, 512  

\bibitem[Evans $\&$ Read(1998)]{wynjenny}
Evans, N. W. and Read, J. C. A., 1998, \mnras, 300, 83.

\bibitem[Evans(1994)]{wynpower}
Evans, N. W., 1994, \mnras, 267, 333.

\bibitem[Famaey et. al(2005)]{famaey05}
Famaey, B., Jorissen, A., Luri, X., Mayor, M., Udry, S., Dejonghe, H. and Turon, C., 2005, \aa, 430, 165.

\bibitem[Famaey et. l(2005)]{famaey_conf}
Famaey, B., Jorissen, A., Luri, X., Mayor, M., Udry, S., Dejonghe, H. and Turon, C., 2005, {\em The Three-Dimensional Universe with Gaia}, eds. Turon, C., O'Flaherty, K. and Perryman, M., 129.

\bibitem[Fux(2001)]{fux_aa} 
Fux, R., 2001, \aj, 373, 511.

\bibitem[Fux(2000)]{fux} 
Fux, R., 2000, {\em Galactic Dynamics from the Early Universe to the present}, eds. Combes, F., Mamon, G.A. and Charmandaris, V., ASP Conference Series Vol 197, 27.

\bibitem[Helmi et. al(1999)]{helmi}
Helmi, A., White, S. D. M., de Zeeuw, P. T. and Zhao, H., 1999, \nat, 402, 53.

\bibitem[Johnston et. al(2001)]{johnston01}
Johnston, S., Koribalski, B., Weisberg, J. M. and Wilson, W., 2001, \mnras, 322, 715.

\bibitem[Kalnajs(1991)]{agris}
A. J. Kalnajs, 1991, {\em Dynamics of Disk Galaxies}, eds. B. Sundelius, 323.

\bibitem[Loredo(1992)]{loredo}
Loredo, T. J., 1992, {\em Statistical Challenges in Modern Astronomy}, eds. Feigelson, E. D. and Babu, G. J., Spirnger-Verlag, New York, 275, http://www.astro.cornell.edu/staff/loredo/bayes/promise.pdf. 

\bibitem[Lynden-Bell $\&$ Kalnajs(1972)]{donaldagris}
Lynden-Bell, D. and Kalnajs, A., 1972, 157, 1.

\bibitem[Melnik(2006)]{melnik}
Melnik, A., 2006, {\em Astron. Lett.}, 32, 7.

\bibitem[Mestel(1963)]{leon}
Mestel, L., 1963, \mnras, 126, 553.

\bibitem[Navarro et. al(2004)]{navarro04}
Navarro, J. F., Helmi, A. and Freeman, K. C., 2004, \apj, 601, L43.

\bibitem[Palous $\&$ Hauck(1986)]{palous}
Palous, J. and Hauck, B., 1986, 162, 54.

\bibitem[Quillen \& Minchev(2005)]{quillenminchev} 
Quillen, A.~C., \& Minchev, I.\ 2005, \aj, 130, 576. 

\bibitem[Quillen(2003)]{quillen}
Quillen, A. C., 2003, \aj, 125, 785.

\bibitem[Raboud et. al(1998)]{raboud}
Raboud, D., Grenon, M., Martinet, L., Fux, R. and Udry, S., 1998, \aa, 336L, 61.

\bibitem[Rautiainen $\&$ Salo(1999)]{rautiainen}
Rautiainen, P. and Salo, H., 1999, \aa, 348, 737.

\bibitem[Saha(1998)]{prasenjitbook}
Saha, P., 1998, {\em Principles of Data Analysis}, Capella Archive, www-theorie.physik.unizh.ch/~psaha/pda.

\bibitem[Sellwood $\&$ Binney(2002)]{binneysellwood}
Sellwood, J. A. and Binney, J., 2002, \mnras, 336, 785.


\bibitem[Skuljan et. al(1999)]{skuljan}
Skuljan, J., Hearnshaw, J. B. and Cottrell, P. L., 1999, \mnras, 308, 731.

\bibitem[Steinmetz et, al(2006)]{steinmetz_rave}
M. Steinmetz, T. Zwitter, A. Siebert, F.G. Watson, K.C. Freeman, U. Munari, R. Campbell, M. Williams, G.M. Seabroke, R.F.G. Wyse, Q.A. Parker, O. Bienayme, S. Roeser, B.K. Gibson, G. Gilmore, E.K. Grebel, A. Helmi, J.F. Navarro, D. Burton, C.J.P. Cass, J.A. Dawe, K. Fiegert, M. Hartley, K.S. Russell, W. Saunders, H. Enke, J. Bailin, J. Binney, J. Bland-Hawthorn, C. Boeche, W. Dehnen, D.J. Eisenstein, N.W. Evans, M. Fiorucci, J.P. Fulbright, O. Gerhard, U. Jauregi, A. Kelz, L. Mijovic, I. Minchev, G. Parmentier, J. Penarrubia, A.C. Quillen, M.A. Read, G. Ruchti, R.-D. Scholz, A. Siviero, M.C. Smith, R. Sordo, L. Veltz, S. Vidrih, R. von Berlepsch, B.J. Boyle and E. Schilbach, 2006, \aj, 132, 1645.


\bibitem[Vallee(2002)]{vallee02}
Vallee, Jacques P., 2002, \apj, 566, 261.


\bibitem[Walker $\&$ Ford(1969)]{fordwalker}
Walker, G. and Ford, J., 1969, {\em Phys. Rev.}, 188, 416.

\bibitem[Woolley(1961)]{woolley}
Woolley, R., 1961, {\em The Observatory}, 81, 203. 


\end{thebibliography}

\end{document}